\documentclass[12pt,preprint]{aastex}

\usepackage{graphicx}
\usepackage{txfonts}

\newcommand{\hi}{\ion{H}{1}}
\newcommand{\nai}{\ion{Na}{1}}
\newcommand{\li}{\ion{Li}{1}}
\newcommand{\mgi}{\ion{Mg}{1}}
\newcommand{\fei}{\ion{Fe}{1}}
\newcommand{\feii}{\ion{Fe}{2}}

\newcommand{\hei}{\ion{He}{1}}
\newcommand{\caii}{\ion{Ca}{2}}
\newcommand{\oi}{[\ion{O}{1}]}
\newcommand{\oxy}{\ion{O}{1}}

\newcommand{\msun}{M$_{\odot}$}
\newcommand{\rsun}{R$_{\odot}$}
\newcommand{\msunyr}{M$_{\odot}$\,yr$^{-1}$}
\newcommand{\macc}{$\dot{M}_{acc}$}

\newcommand{\lacc}{$L_{\mathrm{acc}}$}
\newcommand{\rstar}{$R_{\mathrm{*}}$}

\newcommand{\mstar}{$M_{\mathrm{*}}$}

\slugcomment{To appear in Ap. J.}

\begin{document}

\title{The 2015-2016 outburst of the classical EXor V1118 Ori}

\author{
T. Giannini\altaffilmark{1}, S. Antoniucci\altaffilmark{1}, D. Lorenzetti\altaffilmark{1}, U. Munari\altaffilmark{2}, G. Li Causi\altaffilmark{1,3}, C. F. Manara\altaffilmark{4}, B. Nisini\altaffilmark{1}, A. A. Arkharov\altaffilmark{5}, S. Dallaporta\altaffilmark{6}, A. Di Paola\altaffilmark{1}, A. Giunta\altaffilmark{1}, A. Harutyunyan\altaffilmark{7},  S.A. Klimanov\altaffilmark{5}, A. Marchetti\altaffilmark{8},  G.L. Righetti\altaffilmark{6}, A. Rossi\altaffilmark{9}, F. Strafella\altaffilmark{10}, V. Testa\altaffilmark{1}
}
\altaffiltext{1}{INAF - Osservatorio Astronomico di Roma, via Frascati 33, 00078, Monte Porzio Catone, Italy, teresa.giannini@oa-roma.inaf.it}
\altaffiltext{2}{INAF - Osservatorio Astronomico di Padova, via dell' Osservatorio 8, 36012, Asiago (VI), Italy}
\altaffiltext{3}{INAF -Istituto di Astrofisica e Planetologia Spaziali,  via Fosso del Cavaliere 100, 00133, Roma, Italy}
\altaffiltext{4}{Scientific Support Office, Directorate of Science, European Space Research and Technology Centre (ESA/ESTEC), Keplerlaan 1, 2201, AZ Noordwijk, The Netherlands}
\altaffiltext{5}{Central Astronomical Observatory of Pulkovo, Pulkovskoe shosse 65, 196140, St.Petersburg, Russia}
\altaffiltext{6}{ANS Collaboration, Astronomical Observatory, 36012, Asiago (VI), Italy}
\altaffiltext{7}{Fundaci\'{o}n Galileo Galilei – INAF, Telescopio Nazionale Galileo, 38700, Santa Cruz de la Palma, Tenerife, Spain}
\altaffiltext{8}{INAF - Osservatorio Astronomico di Brera, Via Brera 28, 20122, Milano, Italy}
\altaffiltext{9}{INAF - Osservatorio Astronomico di Bologna, via Ranzani 1, 40127, Bologna, Italy}
\altaffiltext{10}{Dipartimento di Matematica e Fisica, Universit\'{a} del Salento, 73100 Lecce, Italy}

\begin{abstract}
After a quiescence period of about 10 years,the classical EXor source V1118 Ori has undergone an accretion
outburst in 2015 September. The maximum brightness (DV $\ga$ 4 mag) was reached in 2015 December and was
maintained for several months. Since 2016 September, the source is in a declining phase. Photometry and low/
high-resolution spectroscopy were obtained with
MODS and LUCI2 at the {\it Large Binocular Telescope}, with the facilities at the Asiago 1.22 and 1.82 m  telescopes, and with GIANO at the {\it Telescopio Nazionale Galileo}. The spectra are dominated by  emission lines of \hi\ and neutral metallic species. From line and continuum analysis we derive the mass accretion rate and its evolution during the outburst. Considering that extinction may vary between 1.5 and 2.9 mag, we obtain \macc\,=\,0.3$-$2.0 10$^{-8}$ \msunyr\, in quiescence and \macc\,=\,0.2$-$1.9 10$^{-6}$ \msunyr\, at the outburst peak.
The Balmer decrement shape has been interpreted by means of line excitation models, finding that from quiescence to outburst peak, the electron density has increased from  $\sim$ 2 10$^9$ cm$^{-3}$ to $\sim$ 4 10$^{11}$ cm$^{-3}$. The profiles of the metallic lines are symmetric and narrower than 100 km s$^{-1}$, while \hi\, and \hei\,\,lines show prominent wings extending up to $\pm$ 500 km s$^{-1}$. The metallic lines likely originate at the base of the accretion columns, where neutrals are efficiently shielded against the ionizing photons, while faster ionized gas is closer to the star. Outflowing activity is testified by the detection of a variable P Cyg-like profile of the H$\alpha$ and \hei\, 1.08\,$\mu$m lines.
\end{abstract}


\section{Introduction\label{sec:sec1}}
EXors are pre-main sequence objects showing eruptive variability that is caused by intermittent events of magnetospheric accretion (Shu et al. 1994). Because of its relevance in the overall star formation process, the unsteady mass accretion phenomenon
is a theme that has been largely investigated in the past decade and a comprehensive review of the phenomenological aspects has recently  been given by Audard et al. (2014). What is observed is essentially a sequence
of short-duration outbursts (typically months) occurring at
different timescales (years) and showing different amplitudes. EXors
are believed to share the same triggering mechanism with as FUors objects (Hartmann \& Kenyon 1985), but they present substantial differences such as
shorter and more frequent outbursts, spectra dominated by emission lines instead of absorption lines, and lower values of the mass accretion rate. 
Since a detailed model of the disk structure and its evolution does not yet exist for EXor stars, their phenomenology has so far been interpreted by adopting the theoretical
approaches that have been developed for studying FUor events (e.g. Zhu et al. 2009). In this framework  D'Angelo \& Spruit (2010) provided quantitative predictions for the episodic accretion of piled-up material at the inner edge of the disk.\

Although the details of the mechanism responsible for the onset of EXor accretion outbursts are not known, two main scenarios have been proposed, which involve (1) disk instability, and (2) perturbation of the disk by an external body. The first group of models considers gravitational, thermal, or magnetospheric instabilities.
While gravitational instabilities (e.g. Adams \& Lin 1993) do not predict observed short time-scale of the photometric fluctuation, thermal instabilities (Bell $\&$ Lin 1994) offer a more acceptable explanation. They occur when the disk temperature reaches a value of about 5000 K and the opacity becomes a strong function of any (even very small) temperature fluctuation. Thermal models, however, present some difficulties and limitations that are mainly related to the unrealistic constraints on the disk viscosity. Alternatively, outbursts might
be indirectly triggered by stellar dynamo cycles via radial diffusion of the magnetic field across the disk (Armitage 2016).
In the second group of models, several relevant mechanisms have been considered, such as a massive planet that opens up a gap in the disk (Lodato $\&$ Clarke 2004) that is then able to trigger thermal instabilities at the gap itself. Alternatively, accretion bursts can be caused by an external trigger such as a close encounter in a binary system (Bonnell \& Bastien 1992; Reipurth \& Aspin 2004).\ 

The lack of specific models is also related to a description of the onset phase that is not detailed enough. Hence a
monitoring that continuously follows the pre-outburst and the outburst  evolution is fundamental in order to provide the variations of both the photometric 
(light curve, colors) and spectroscopic properties (line excitation, ionization, dynamics). Remarkable efforts in
this sense have been made by Sicilia-Aguilar et al. (2012) and Hillenbrand et al. (2013), who give details of the last outburst of EX Lup, the prototype of the class, and the more embedded V2492 Cyg.
Another classical EXor frequently monitored is V1118 Ori ($\alpha_{J2000.0}$= 05$^{h}$ 34$^{m}$ 44${\farcs}$745, $\delta_{J2000.0}$=
$-$05$^{\circ}$ 33$^{\prime}$ 42${\farcs}$18): its flares-ups in the past 40 years have been photometrically documented in the optical bands (e.g. Parsamian et al. 1993, 1996, 2002; Garcia Garcia \& Parsamian 2000, 2008; Audard et al. 2005, 2010; Jurdana-{\v S}epi{\'c} et al. 2017),
and in particular, we followed different phases of activity during the past 10 years (Lorenzetti et al. 2006, 2009, 2015a, hereafter Paper\,I), while 
Audard et al. have performed a detailed photometric monitoring of the 2005 outburst that spanned a wide interval of frequencies from  near-
and mid-IR to X-rays. In the framework of our monitoring program EXORCISM (EXOR optiCal and Infrared Systematic Monitoring, Antoniucci et al. 2014), we have discovered
a new outburst of V1118 Ori in September 2015 (Lorenzetti et al. 2015b, Giannini et al. 2016, hereafter Paper\,II). 

In this work we present the optical and near-IR follow-up observations of this last outburst. Our aim is to provide a global picture of all the outburst phases (rising, peak, and declining) from an observational point of view. Photometric and spectroscopic observations are presented in Section\,\ref{sec:sec2}, and are analyzed and discussed in Section\,\ref{sec:sec3}. 
Our concluding remarks are given in Section \,\ref{sec:sec4}.

\section{Data\label{sec:sec2}}

\subsection{Literature and archival photometry\label{sec:sec2.1}}
\begin{figure}
\epsscale{1.0}
\plotone{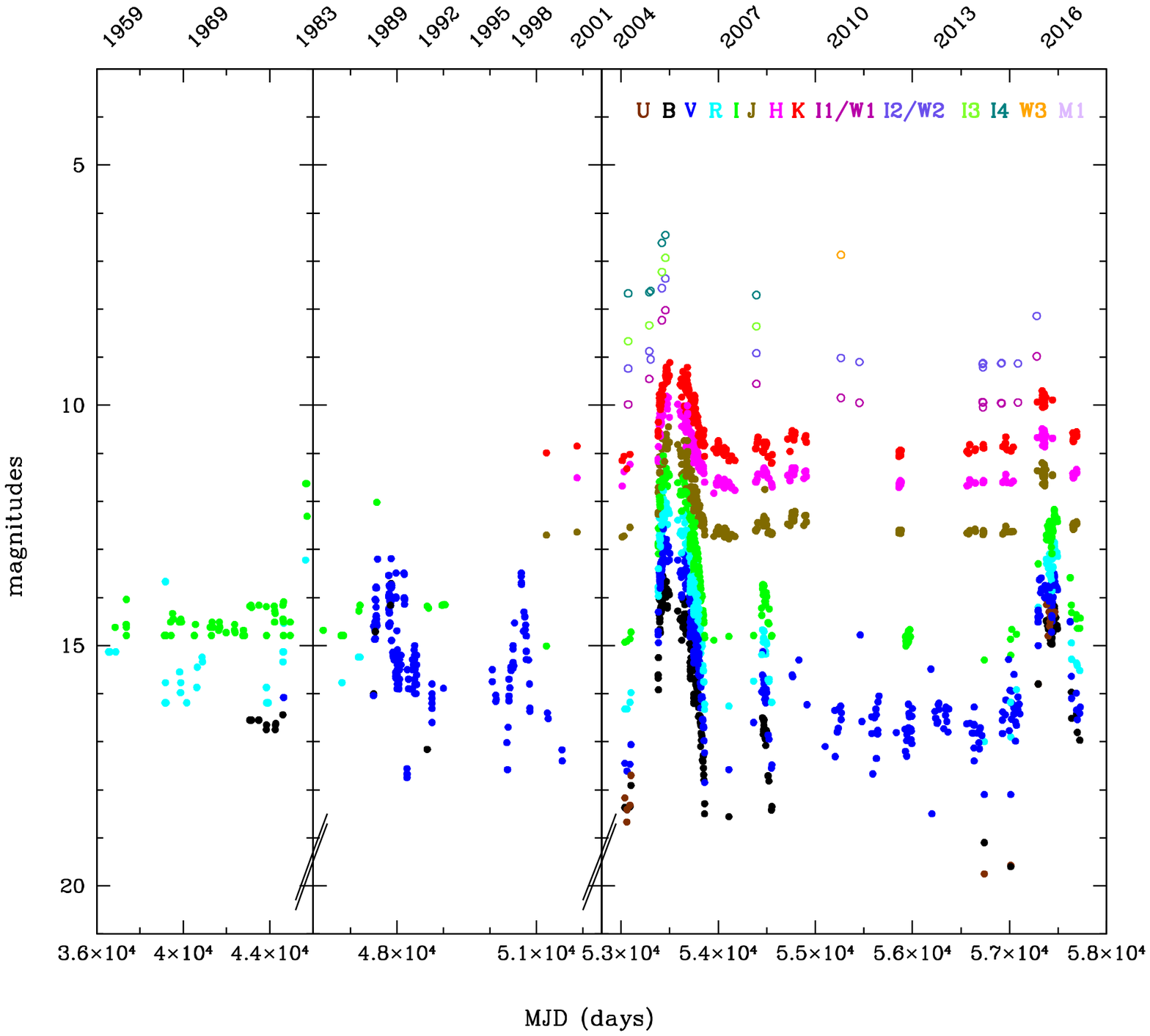}
\caption{Historical light curve of V1118 Ori. Optical/near-IR and mid-IR magnitudes are represented with filled and open circles, respectively. Magnitudes in different filters are depicted with different colors, as indicated in the top-right label.
$UBVRIJHK$ photometry was taken from: Audard et al. 2005, 2010; Garcia Garcia et al. 1995, 2006; Garcia Garcia \& Parsamian 2000, 2008;
Gasparian \& Ohanian 1989; Jurdana-\v{S}epi\'c et al. 2017; Hayakawa et al. 1998; Hillenbrand 1997; Parsamian \& Gasparian 1987; Parsamian et al. 1993, 1996, 2002; Papers\,I, II; Verdenet et al. 1990; Williams et al. 2005. We have also retrived data from the catalogs: AAVSO ($BVRI$), 2MASS ($JHK$), DENIS ($IJK$), WISE (3.4\,/\,4.6\,/\,12 $\mu$m, W1-W3), Spitzer/IRAC (3.6\,/\,4.5\,/\,5.8\,/8.0\,$\mu$m, I1-I4), and Spitzer/MIPS (24\,$\mu$m, M1). \label{fig:fig1}}
\end{figure}

To have a complete view of the photometric activity of V1118 Ori, we have searched the literature and public archives to construct its multiwavelength historical light curve. This is depicted in Figure\,\ref{fig:fig1}  (where our new data are also shown) and commented in Section\,\ref{sec:sec3.1.1}.
We collected data since 1959 for the bands $UBVRIJHK$ (references are in the Figure caption), {\it WISE}/1-3 (at 3.4\,/\,4.6\,/\,12\,$\mu$m),
{\it Spitzer}/IRAC (3.6\,/\,4.5\,/\,5.8\,/8.0\,$\mu$m), and {\it Spitzer}/MIPS (24\,$\mu$m). {\it Spitzer} data are published in Audard et al. (2010), while {\it Wide-field Infrared Survey Explorer}  (WISE) data are reported here for the first time (Sect.\,\ref{sec:sec2.2.3}).
For completeness, we also  account for a few data before 1960s, which indicate a remarkable level of activity  ($m_{\rm{pg}}$ $\sim$ 14.0) in 1939, 1956, and 1961 (Paul et al. 1995).

\subsection{New photometry\label{sec:sec2.2}}
The new photometry covers the period 2015 December-2016 December. Together with the photometry reported in Papers\,I, II it follows the evolution 
from quiescence to the post-outburst phases. In Figure 2 we
show the light curve from 2015 January-2016 December,
where we have empirically identified five phases to which we
refer in the analysis: (1) quiescence: up to 2015 March; (2)
rising: 2015 October; (3) peak: 2015 November-2016 April;
(4) declining: 2016 September; and (5) post-outburst: since
2016 December.
  
\subsubsection{Optical photometry\label{sec:sec2.2.1}}
$BVR_{\rm C}I_{\rm C}$ optical photometry of V1118 Ori has been
obtained with the {\it Asiago Novae and Symbiotic stars} (ANS) Collaboration telescopes 36 and 157, 0.3m $f$/10 instruments
located in Cembra and Granarolo (Italy).  Technical details and operational
procedures of the ANS Collaboration network of telescopes are presented by
Munari et al.  (2012), while analysis of the photometric performances and
multi-epoch measurements of the actual transmission profiles for all the
photometric filter sets are discussed by Munari \&
Moretti (2012).  The same local photometric sequence was used at both
telescopes in all observing epochs, ensuring a high consistency of the data. 
The sequence was calibrated from APASS survey data (Henden et al.  2012,
Henden \& Munari 2014) using the SLOAN-Landolt transformation equations
calibrated in Munari (2012) and Munari et al.  (2014a,b).
Reduction was achieved  by using standard procedures for bad-pixel cleaning, bias and dark removal, and flat fielding.
All
measurements were carried out with aperture photometry, the long focal
length of the telescopes and the absence of nearby contaminating stars not
requiring to revert to PSF-fitting. 

\begin{figure}
\epsscale{1}
\plotone{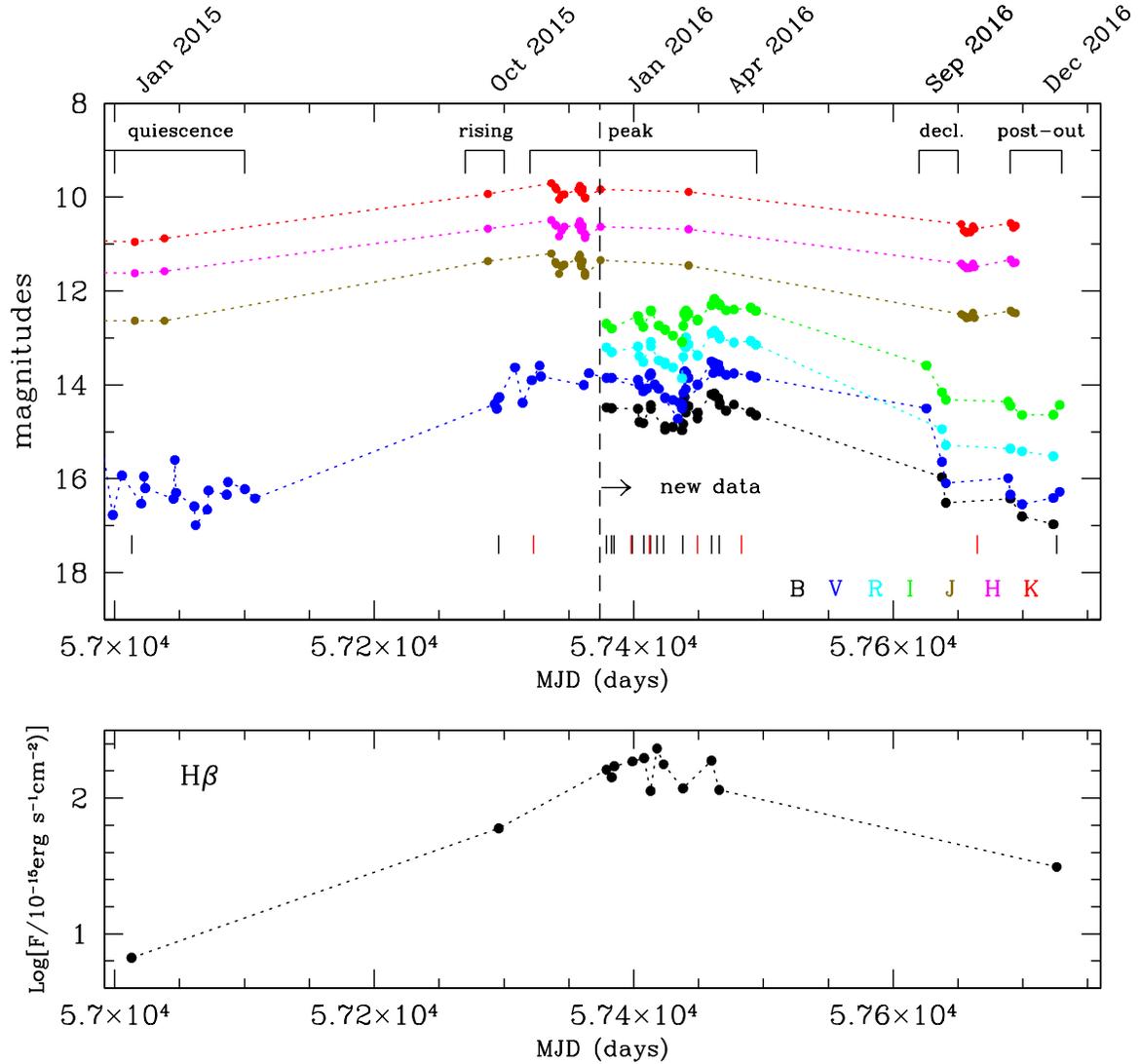}
\caption{Top panel: optical and near-IR light curve in the period 2015 January-2016 December. Colors indicate the same bands as in Figure\,\ref{fig:fig1}. Data taken  after the date marked with a dashed vertical line are reported in this paper for the first time, while those before that date are taken from the literature. These are shown to have a global view of the 2015-2016 outburst. Different temporal phases are labeled: (1) quiescence:  up to 2015 March; (2) rising: 2015 October; /3) peak: 2015 November-2016 April; (4) declining: 2016 September; and (5) post-outburst: since 2016 December.  
Vertical black  and red segments indicate the dates when optical and near-IR spectra were taken. Bottom panel: flux of the H$\beta$ line as a function of time.\label{fig:fig2}}
\end{figure}

The $BV$$R_{\rm C}$$I_{\rm C}$ photometry of V1118 Ori is given in Table~\ref{tab:tab1}
and shown in Figure\,\ref{fig:fig2}, top panel. 
The source has remained bright (with daily variations of up to 0.4 mag in $V$) throughout the monitoring period until 2016 April, when observations had to be momentarily alted for Solar conjunction. They restarted in 2016 August, and at that time the visual magnitude dropped by about 1 mag. Since 2016 August, V1118 Ori  has continued to rapidly fade to $V\approx$ 16.5 mag, even if it has not reached the magnitude ($V\approx$ 17$-$18 mag) typical of its quiescent state at present.

\subsubsection{Near-infrared photometry\label{sec:sec2.2.2}}

$JHK$ aperture photometry was achieved with the 1.1m AZT-24 telescope at Campo Imperatore  (L'Aquila, Italy) with the
SWIRCAM camera (D'Alessio et al. 2000). 
Data reduction was performed  by using standard procedures for bad-pixel cleaning, flat fielding, and sky subtraction. Calibration was achieved on the basis of $2MASS$ photometry of 
several bright stars in the field. We present in Table\,\ref{tab:tab2} and Figure\,\ref{fig:fig2} (top panel) the data obtained between  2015 December
and  2016 November, which follow those provided in Papers\,I and II. The peak of the infrared activity occurred between 2015 December and 2016 February. No data are available between 2016 March and September; after this period, the source has faded and reached the quiescence value in all the near-IR bands (see Paper\,I).

\subsubsection{{\it WISE} photometry\label{sec:sec2.2.3}}

V1118 Ori was also observed in its quiescent phase by WISE, (Wright et al. 2010) during both the cryogenic  (2009 December 2009 - 2010 August) and the 3-band (2010 August - September) surveys (see Table\,\ref{tab:tab3}). Average values of the {\it WISE}  magnitudes are [3.4]\,=\,9.9, [4.6]\,=\,9.0, [12]\,=\,6.9, and [22]\,$>$\,3.4 mag. More recent  observations have been collected in the first two bands during the ongoing {\it Neowise-R} survey.  Measurements taken on  2015 September 16 are [3.4]\,=\,8.9 and [4.6]\,=\,8.1, with a brightness increase of $\approx$ 1 mag  in both bands with respect to the quiescence values.

\subsection{Spectroscopy\label{sec:sec2.3}}

V1118 Ori has been spectroscopically observed in low- and high-resolution configurations at optical and near-IR wavelengths. A journal of the spectroscopic observations is given in Table\,\ref{tab:tab4}. 

\subsubsection{Optical spectroscopy\label{sec:sec2.3.1}}
Four optical low-resolution spectra have been collected on 2016 January
 12, 26, and 31 and 2016 December 4 with the 8.4\,m Large Binocular Telescope (LBT) using the Multi-Object Double Spectrograph (MODS - Pogge et al. 2010). All observations have been taken with the dual grating mode (Blue + Red channels) and integrated 1500 s in the spectral range 3200$-$9500~\AA\, by using a 0$\farcs$8 slit 
 ($\Re \sim$ 1500). Data reduction was performed at the Italian LBT Spectroscopic Reduction Center\footnote{http://www.iasf-milano.inaf.it/Research/lbt\_rg.html} by means of scripts optimized for LBT data.
Steps of the data reduction of each two-dimensional spectral image are correction for dark and bias, bad-pixel mapping, flat-fielding, sky background subtraction, and extraction of one-dimensional spectrum by integrating the stellar trace along the spatial direction. Wavelength calibration was obtained from the spectra of arc lamps, while flux calibration was achieved using the ANS Collaboration telescopes photometry, taken within one day from the spectroscopic observations.
 
Seven additional spectra  (3300$-$8050~\AA)  at $\Re$ $\sim$ 2400 were obtained with the 1.22m
telescope + B\&C spectrograph operated in Asiago by the University of Padova.  
The slit has been kept fixed at a width of 2$\arcsec$ and was always aligned with the parallactic angle for optimal absolute
flux calibration.  
 
Finally, six high-resolution spectra in the range 3600$-$7300~\AA\, were taken with the REOSC Echelle
spectrograph mounted on the 1.82m telescope operated in Asiago by the
National Institute of Astrophysics (INAF).  The spectrograph 
covers
the whole wavelength interval in 32 orders without
interorder gaps.  The slit width was set to provide a resolving
power of 20\,000, and - as for the 1.22m - the slit was always aligned with
the parallactic angle for optimal absolute flux calibration.
 
The spectra from the Atwo siago telescopes were reduced within IRAF\footnote{IRAF is distributed by the National Optical Astronomy Observatory, which is operated by the Association of the Universities for Research in Astronomy, inc. (AURA) under cooperative agreement with the National Science Foundation.}  following 
all the steps cited above for the reduction of the MODS spectra. Flux calibration was achieved by means of 
standards observed at similar airmass before and after V1118 Ori.  While this
ensures an excellent correction of the instrumental response, the zero-point
of the flux scale was sometimes affected by the unstable sky
transparency.  Check (and correction whenever necessary, always $\leq$30\%) of
the zero-point has been performed by integrating on the spectrum the flux
through the $B$,$V$ and $R$-band profiles and comparing to values from the photometric
campaign.

\begin{figure}
\epsscale{1.0}
\plotone{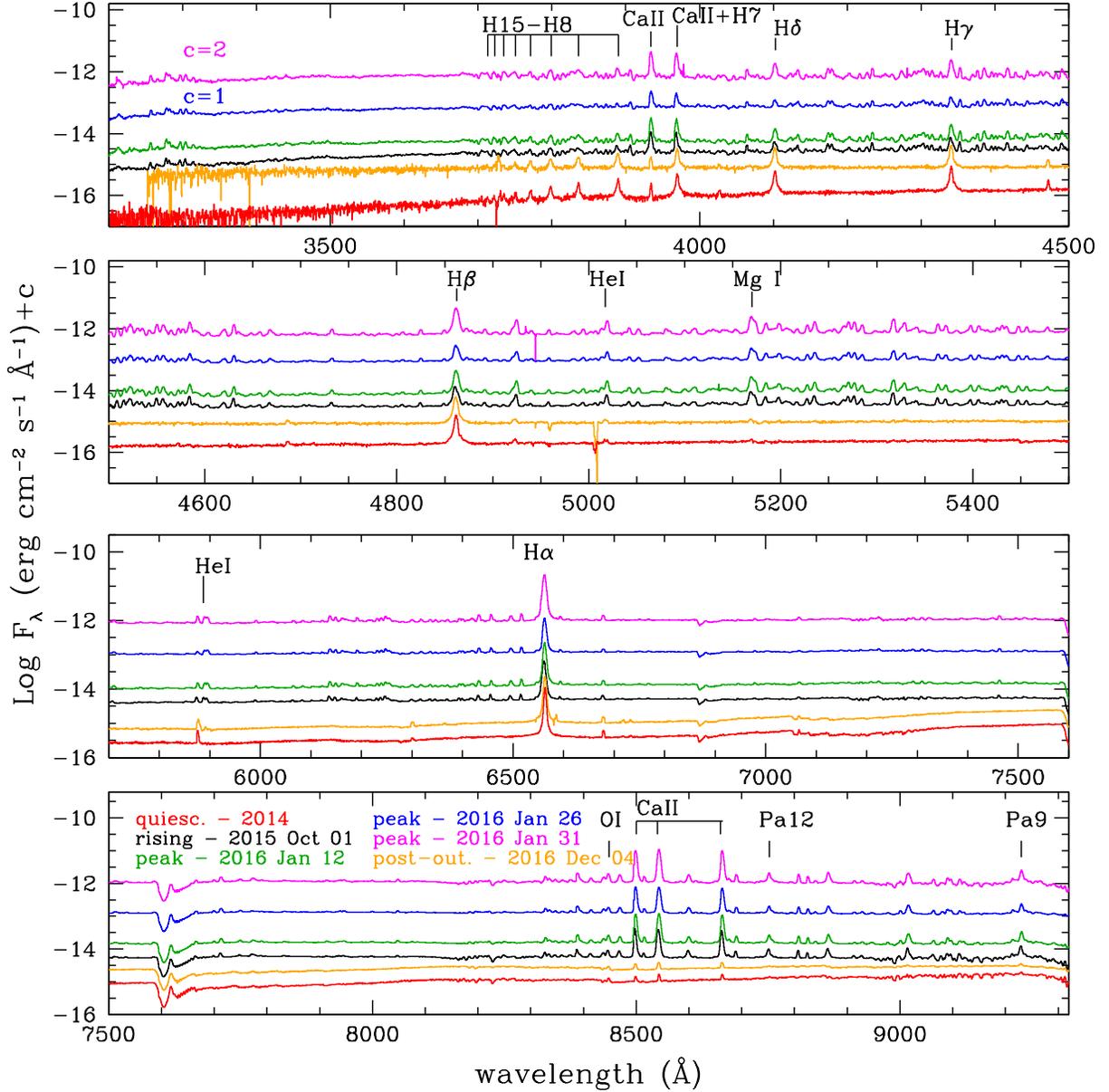}
\caption{Optical (LBT/MODS) spectra of V1118 Ori taken in 2016 January (peak: green, blue, and pink) and 2016 December (post-outburst: orange), shown in comparison with the spectra taken 
in quiescence (red, Paper\,I) and during the rising phase (black, Paper\,II). For a better visualization a constant (indicated in the top panel) was added to the spectra taken on January 26 and 31. Main emission lines are labeled.\label{fig:fig3}}
\end{figure}

In Figure\,\ref{fig:fig3} the four MODS spectra are depicted in comparison with the quiescence spectrum  (Paper\,I) and that obtained during the rising phase  (Paper\ II). The same forest of lines as observed in the 2015 October spectrum is present in the 
burst spectra, but further increased in brightness. The large majority are recombination lines or lines of neutral and ionized metals (\hi, \hei, \caii, \fei, and \feii), most of them blended with each other. 
 We give in Table\,\ref{tab:tab5} the line fluxes of bright and unblended lines that are used in the spectral analysis. 
Their flux is computed by integrating  the signal below the spectral profile. The FWHM is compatible with the instrumental one because the lines are not resolved in velocity at the adopted resolution. The associated uncertainty is estimated by evaluating the {\it rms} in a wavelength range close to the line, and then multiplying it by the FWHM. 

\begin{figure}
\epsscale{1.0}
\plotone{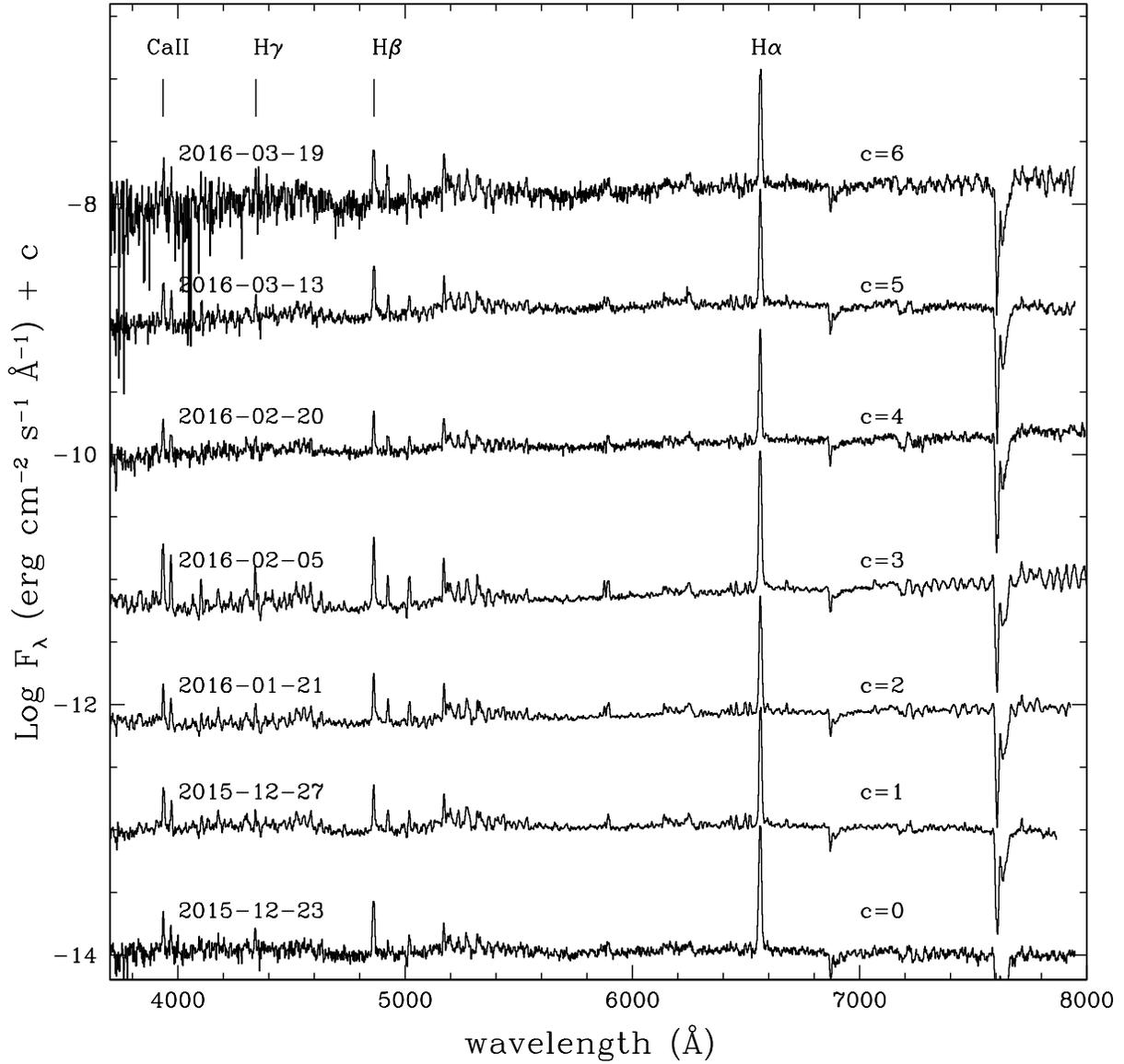}
\caption{Optical low-resolution spectra of V1118 Ori taken with the 1.22m Asiago telescope during the peak phase. For a better visualization a constant  was added to the flux density. Main emission lines are labeled.\label{fig:fig4}}
\end{figure}

The low-resolution Asiago spectra are shown in Figure\,\ref{fig:fig4}. Together with \hi\, and \caii\, lines, other features of metallic lines are prominent in the spectrum, but they result from blends of two or more lines, so their flux is not useful for the line analysis.    

Most of the permitted lines that appear blended in the optical low-resolution spectrum are clearly identified at the spectral resolution achieved in the echelle spectra. A list and comments on these lines are provided in Appendix\,A and in Section\,\ref{sec:sec3.2.4}, respectively. Profiles of the most interesting lines (in particular the Balmer lines) are presented in Section\,\ref{sec:sec3.3}.

Fluxes of unblended bright lines observed with the Asiago telescopes are listed in Table\,\ref{tab:tab6}. The sensitivity limit is about a factor of ten lower than the limit of MODS, therefore reliable fluxes can be given only for some permitted lines (not reported in Table\,\ref{tab:tab6}), along with the Balmer and \caii\,\,H lines.  
Whenever both a low- and a high-resolution spectrum is available at a certain date, the fluxes agree each other within 20\%, with the exception of the flux of H$\alpha$, which shows prominent wings that are better resolved at higher dispersion. In Table\,\ref{tab:tab6} we therefore give the average of the two determinations, except for  H$\alpha$, for which we report the flux measured in the high-resolution spectrum.
 
Finally, we note that many lines present a remarkable level of variability on timescales of days (see Tables\,\ref{tab:tab5} and \ref{tab:tab6}). For example,  H$\alpha$ and H$\beta$  fluxes increase by about a factor of two from January 26 to 31. As we show as an example in the bottom panel of Figure\,\ref{fig:fig2}, the variation of F(H$\beta$) roughly follows that of the continuum, although with a certain delay. In particular, the continuum fades faster than lines. This is evident for the last spectroscopic point  (MJD 57726), when the continuum level was close to that of quiescence (top panel), while F(H$\beta$) was still about a factor of 5 brighter. This circumstance appears to be a common feature of EXor variability (Paper\,I).

\subsubsection{Near-IR spectroscopy\label{sec:sec2.3.2}}
Three near-infrared, low-resolution spectra have been obtained with the LUCI2 instrument at LBT on  2016 January 25,  March 2, and October 4.
The observations were carried out with the G200 low-resolution grating coupled with the 1$\farcs$0 slit on the first two dates and with the 0$\farcs$75 slit on October 4. The standard ABB'A' technique was adopted to perform the observations using the $zJ$ and $HK$ grisms, for a total integration time of 12 and 8 minutes, respectively. The final spectrum covers the wavelength range 1.0$-$2.4 $\mu$m at $\Re \sim$1000. Data reduction was performed at the Italian LBT Spectroscopic Reduction Center.
The raw spectral images were flat-fielded, sky-subtracted, and corrected for optical distortions
in both the spatial and spectral directions. Telluric absorptions were removed using the normalized spectrum of a telluric standard star, after fitting  its intrinsic spectral features.
Wavelength calibration was obtained from arc lamps, while flux calibration was achieved through the observations of spectrophotometric standards, carried out in the same night as the target. No intercalibration was performed between the $zJ$ and $HK$ parts of the spectrum, since they were optimally aligned. We evaluated possible flux losses by comparing the continuum level at the $JHK$ effective wavelengths (1.25, 1.60, and 2.20 $\mu$m) with the magnitudes of the closest photometric points, obtained within a week from the spectroscopic observations. While for the first two spectra (January 25 and March 2), the agreement is within 20\%, significant flux losses, mainly due to the bad weather conditions during the observation, are registered in the spectrum of October 4. Therefore we used the photometric data of October 1 to calibrate this spectrum, which in any case is much fainter than the first two.
No intercalibration was applied between near-IR and optical spectra as they all were acquired on different dates. 

\begin{figure}
\epsscale{1.0}
\plotone{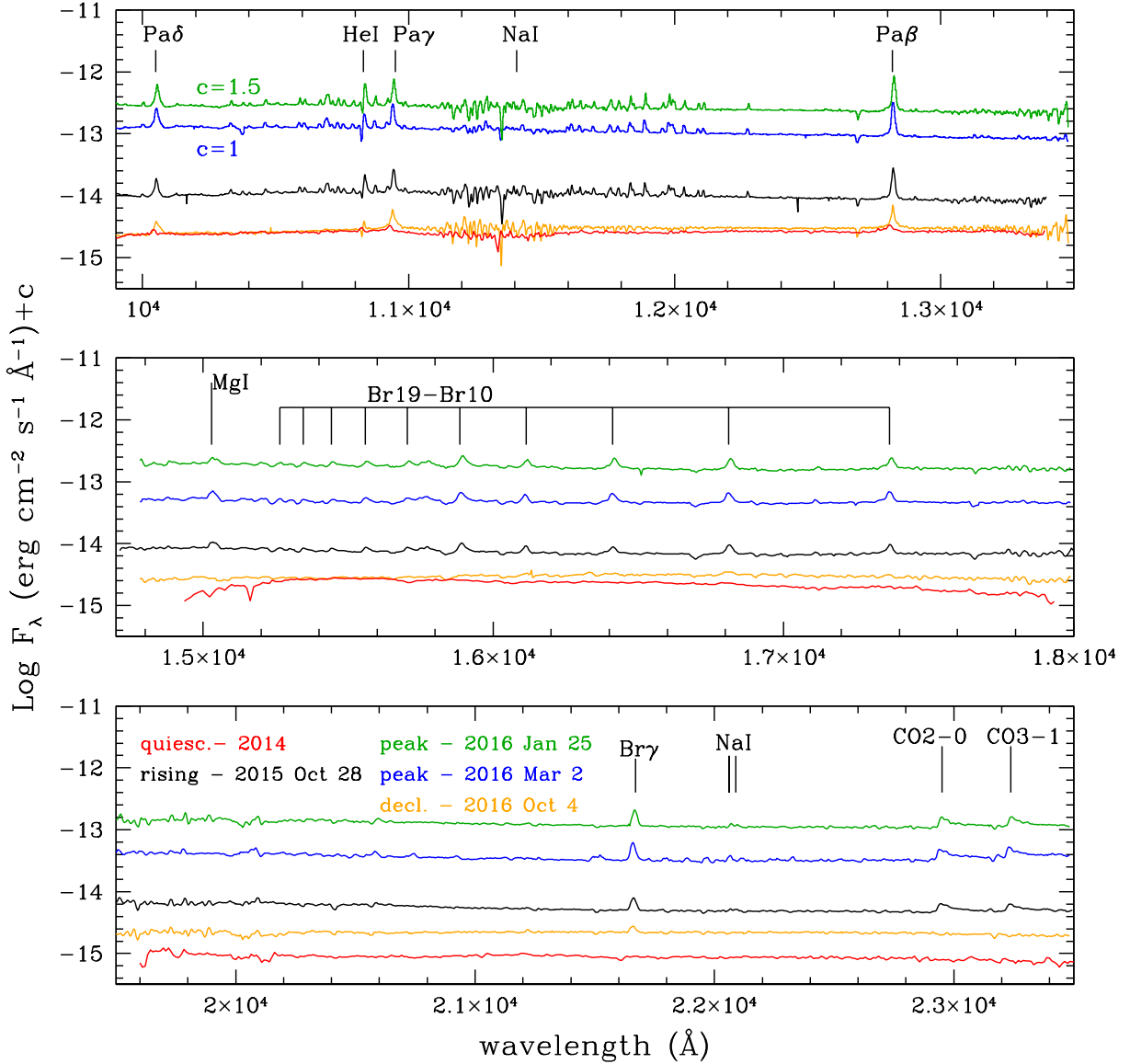}
\caption{Near-IR LUCI2 spectra taken on 2016 January and March (peak: green, blue), and 2016 October (declining: orange) in comparison with spectra taken 
in quiescence (red, Paper\,I) and during the rising phase (black, Paper\,II). For a better visualization a constant (indicated in the top panel) was added to the spectra taken on January 25 and March 2.
Main emission lines are labeled.\label{fig:fig5}}
\end{figure}

The three LUCI2 spectra are shown in Figure\,\ref{fig:fig5}, in comparison with the spectra taken in quiescence  (Paper\,I) and during the rising phase (Paper\ II). Fluxes of the main unblended lines are listed in Table\,\ref{tab:tab7}.
In the peak phase the most prominent lines are \hi\,
 recombination lines of the Paschen and Brackett series, \hei\, 1.08\,$\mu$m, \mgi, and \nai\, lines. CO  2-0 and 3-1 ro-vibrational bands are also observed in emission. Other metallic lines are likely present, but they are too faint or strongly blended to be clearly identified. In the last spectrum, taken during the declining phase, we detect the brightest \hi\, lines,   \hei\, 1.08\,$\mu$m, and H$_2$\,2.12\,$\mu$m. Notably, while \hi\, line fluxes are similar to those during the rising phase (Paper\,II), the continuum level is  closer to the quiescence level. This confirms what we have already noted about the optical lines in Sect.\ref{sec:sec2.3.1}.

Near-IR, high-resolution spectra were obtained on 2016 January 10 and 2016 April 05 with  GIANO at the {\it Telescopio Nazionale Galileo} (La Palma. Spain) during two Director Discretionary Time (DDT) runs. GIANO  provides in a single-exposure cross dispersed echelle spectroscopy at a resolution of 50\,000 ($\delta$v $\sim$ 6 km s$^{-1}$ at $\lambda$\,=\, 1\,$\mu$m) over the 0.95-2.45 $\mu$m spectral range. Since at longer wavelengths the echelle orders become larger than the 2k$\times$2k detector, the spectral coverage in the $K$ band is 75\%, with gaps, for example,  at the Br$\gamma$ and CO 1-0 wavelengths. 
GIANO is fiber-fed with two fibers of 1$\arcsec$  angular diameter at a fixed angular distance of 3$\arcsec$ on sky. The science target is acquired by nodding on fiber, namely
   alternatively acquiring target and sky on fibers A and B, and vice versa. This allows an optimal subtraction of the detector pattern and sky  background.  Data
    reduction  was achieved by means of specific scripts optimized for GIANO\footnote{GIANO$\_$TOOLS, available at the TNG webpage (http://www.tng.iac.es/instruments/giano)} . In particular,
    observations of a uranium-neon lamp were carried out to perform wavelength calibration, while flat-fielding and mapping of the order geometry were achieved through exposures with a tungsten
     calibration lamp. The reduced spectrum is rich in  metallic lines that were barely detected in the low-resolution spectrum, which we list in Appendix\,B and comment in Section\,
      \ref{sec:sec3.2.4}. Profiles of the most interesting lines (\hi\, lines and \hei\, 1.08\, $\mu$m) are presented in Section\,\ref{sec:sec3.3}.

\section{Analysis and Discussion\label{sec:sec3}}

	\subsection{Photometric analysis\label{sec:sec3.1}}
		 
\subsubsection{Light curve\label{sec:sec3.1.1}}

The 2015-2016 outburst is the sixth documented during the recent history (about 40 years) of V1118 Ori; previous events (1983, 1989, 1992, 1997, and 2005) are recognizable in the light curve
in Figure\,\ref{fig:fig1}.
This global view of the V1118 Ori photometric behavior shows
two facts: (1) the present outburst is roughly at the same intensity level ($V_{peak} \sim$ 13.5 mag) and duration ($\sim$ 1 year) as the bursts that occurred between 1989 and 1997, whereas the outburst occurred in 2005 has been the brightest ($V_{peak} \sim$ 12 mag) and maybe the longest ($\ga$ 2 years); (2) in the course of time the monitoring improved progressively in both sampling cadence and spectral coverage.  In particular, systematic
observations have shown that V1118 Ori has remained quiescent for about a decade (from 2006 to 2015). Indeed, four out of the five previously known events have occurred with a roughly
regular cadence of 6-8 years, with the only exception the outburst occurred in 1992 (Garcia Garcia et al. 1995), which was much fainter than all the others ($V\le$ 14.7 mag), however. The evidence that the 2015 outburst is not in phase with the previous outbursts, does not support periodic phenomena  as the cause of the outburst triggering. A quiescence period 
of about 15 years (from 1966 to 1981) has also been shown by a recent analysis of historical plates (Jurdana-{\v S}epi{\'c} et al. 2017).

\begin{figure}
\epsscale{1.0}
\plotone{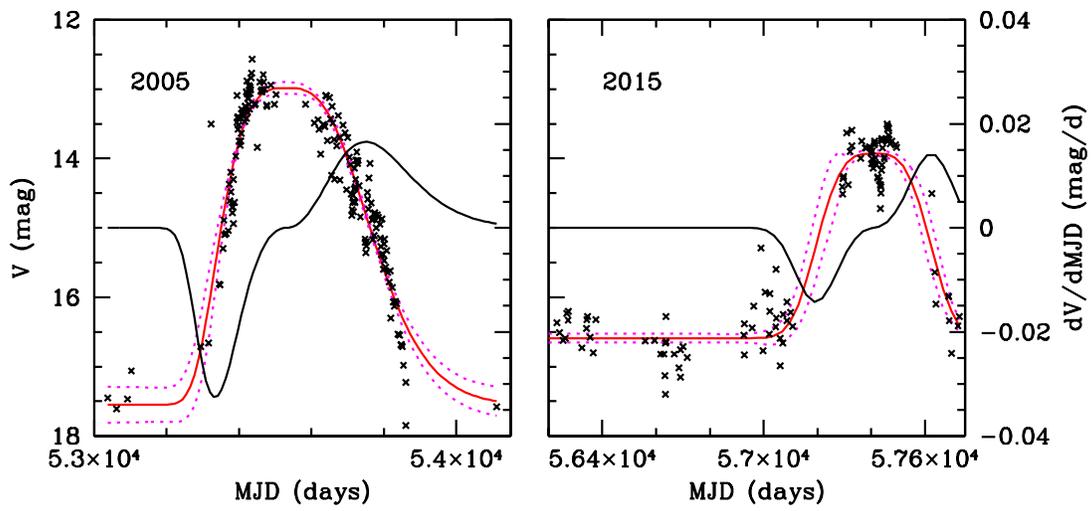}
\caption{\label{fig:fig6} Left panel: $V$ light curve of the 2005 outburst. The red solid curve is the best distorted Gaussian through the data points, whose uncertainty is represented by the pink dotted lines. The black solid line is the derivative of the distorted Gaussian, which represents the velocity in magnitude variation. Right panel: the same as in the left
panel for the 2015 outburst. A large portion of the light curve before the outburst is plotted to show the quiescence level.}
\end{figure}

We have compared the present outburst with the outburst of 2005 to understand whether they have followed a similar evolution in time. 
In Figure\,\ref{fig:fig6} the $V$ light curves of the 2005 and 2015 outbursts are shown in the left and right panel, respectively. While the 2005 outburst has occurred as a sudden
event starting from a very faint level ($V\,\ga$\,17.5 mag), the outburst of 2015 followed a long period that was characterized by a higher average brightness ($<V>$ $\sim$ 16.5 mag) with daily scale fluctuation of several tens of magnitude, presumably unrelated with the accretion burst.
Quantitatively, the two light curves (covering about 1200 days each) have been fitted  with a single distorted Gaussian\footnote{Represented as y(x)$\sim$ exp$^{-f(x,\sigma,s,k)^2}$, with $\sigma$=standard deviation, s=skewness, k=kurtosis.} This is depicted as a red solid curve, while the fit uncertainty is represented
 with pink dotted lines. The derivative to the fitted Gaussian (black line) allows us to account for the speed typical of the different outburst phases. In 2005,
  the rising speed has increased monotonically to values ($\sim$ 0.035 mag/day) about twice the decreasing speed ($\sim$ 0.015 mag/day). Notably, the 2015
   outburst is characterized by a similar rising and declining speed ($\sim$ 0.015 mag/day). This might indicate that while triggering has occurred under different modalities, the relaxing phenomenon has followed as similar evolution.
   
In conclusion, as also confirmed by spectroscopic data, see Sect.\ref{sec:sec3.2.2}, the 2015 outburst appears different from the outburst in 2005, as far as intensity, duration, recurrence time, and rising speed are concerned. Whether the observed behavior is specific of V1118 Ori, or is more in general a common feature of the EXor variability, remains an open question. For the moment, the possible comparison is with the  candidate EXor V2492 Cyg, which shows a slower
rising than declining speed, although the variability of this source is not fully due to disk-accretion phenomena  (Hillenbrand et al. 2013).

\subsubsection{Color analysis\label{sec:sec3.1.2}}

\begin{figure}
\epsscale{1.05}
\plotone{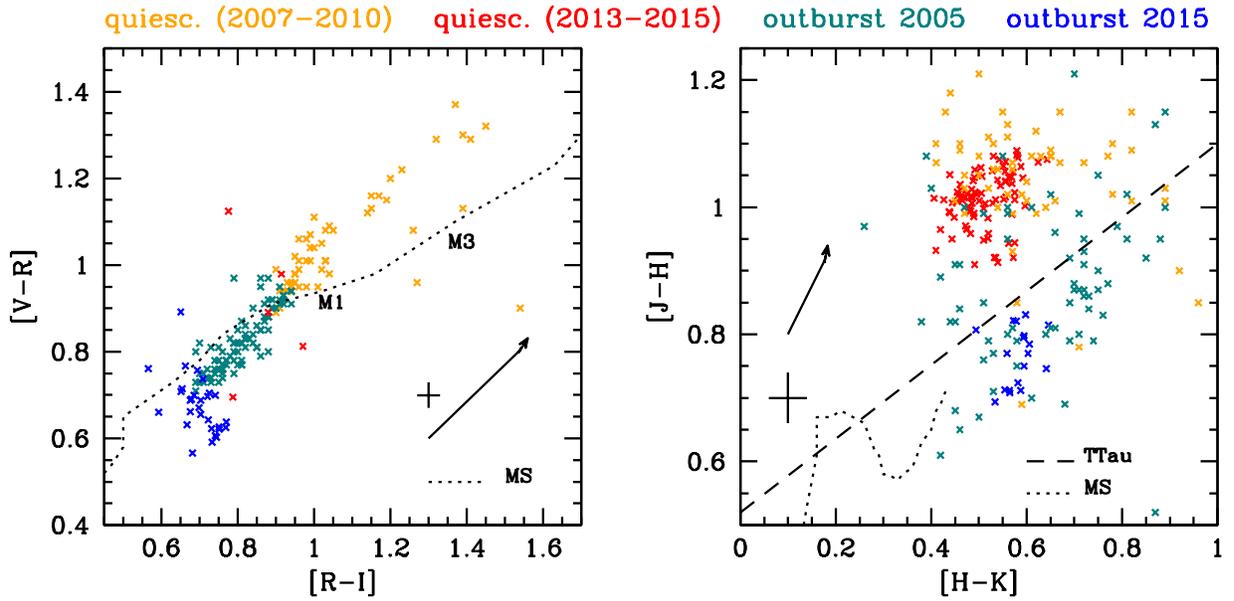}
\caption{ Left panel: optical two-color plot [$V-R$] vs. [$R-I$]. Orange/red crosses: quiescence data ($V \ge$ 15) in the periods 2007-2010/2013-2015; teal-blue/blue crosses: 2005/2015 outbursts data ($V <$ 15). References: present paper; Audard et al. 2010. Dotted line:  unreddened main-sequence. Colors of M1 and M3 stars are indicated. Arrow: extinction vector corresponding to A$_V$\,=\, 1 mag (reddening law of  Cardelli et al. 1989). The average error is indicated with a black cross. 
Right panel: the same as in the left panel for the near-IR two color plot [$J-H$] vs. [$H-K$]. Quiescence and outburst are defined as $H \ge$ 11 and $H <$ 11, respectively.
References: present paper; Papers\,I,\,II; Audard et al. 2010. Dashed line: {\it locus} of the T\,Tauri stars (Meyer et al. 1997).\label{fig:fig7}}
\end{figure}

To investigate the origin of the photometric fluctuation, we plot in Figure\,\ref{fig:fig7}, left panel, the optical [$V-R$] vs. [$R-I$] colors of V1118 Ori measured during quiescence and recent outbursts (2005 and 2015). These are obtained  by combining the data of the present paper with those by Audard et al. (2010). Quiescence and outburst phases are identified by  
$V$ lower or greater, respectively, than 15 mag. The data points of the 2015 outburst follow a similar distribution as those of 2005. As has been shown by Audard et al. (their Figure 11), colors become bluer with increasing brightness, moving almost parallel to the reddening vector in the direction of a decreasing extinction. However, although a variable extinction certainly plays a role (see also Sect.\,\ref{sec:sec3.2.1}),
it cannot be the primary cause of the brightness enhancement typical of outbursts. Indeed, as shown in the Figure, the colors of an {\it unreddened} main-sequence star of the V1118 Ori spectral type (M1-M3) are typically {\it redder} than the outburst colors. Instead, and in agreement with previous studies on EXors (Audard et al. 2010; Lorenzetti et al. 2012), we interpret the blueing of the optical colors
as mainly due to a hotter temperature component appearing during the outburst. Note that in this view the extinction may even {\it increase} during outburst, but not enough to compensate for the blueing of the color indices.  

The near-IR [$J-H$] vs. [$H-K$] color diagram  (Figure\,\ref{fig:fig7}, right panel) is in agreement with the above interpretation. Indeed, although colors of the different phases (quiescence: $H \ge 11$, outburst: $H < 11$)  spread over a large portion of the plot, they roughly segregate into two different portions of the plot, whose connecting direction does not follow that of the reddening vector.

\subsection{Line analysis \label{sec:sec3.2}}

\subsubsection{Extinction estimate\label{sec:sec3.2.1}}
As a first step of the line analysis, we have estimated the extinction toward V1118 Ori. A common and powerful tool to achieve this is to use pairs of optically thin and bright transitions coming from the same upper level, since in this case the difference between observed and theoretical flux ratio depends only on the extinction value. 
We have considered pairs of \hi\, lines whose profiles do not seem affected by evident opacity effects, such as self-absorption or flatted peaks (see Sect.\ref{sec:sec3.3}). These are  Pa$\beta$/H$\gamma$, Pa$\gamma$/H$\delta$, Pa9/H9, Pa12/H12, and Br$\gamma$/Pa$\delta$. Considering all the available spectra, we estimate A$_V$ in the range 1.5$-$2.9 mag, without any significant trend between quiescence and outburst phases. Indeed, extinction fluctuations up to 1 mag are also recognizable in the color-color diagrams of  Figure\, \ref{fig:fig7}.

As an alternative method, we estimate A$_V$ by modeling and fitting the uv/optical MODS continuum (Section \ref{sec:sec3.2.2}). In good agreement with the range found from line ratios, the best fit provides A$_V$\,=\,2.9 mag in quiescence, rising, and at the outburst peak, and  A$_V$\,=\,1.9\, mag in the post-outburst phase.

For comparison, our A$_V$ estimate is also compatible with that  between 1.4 and 1.7 mag  obtained by Audard et al. (2010) on the basis of X-ray observations during the outburst of 2005.
 
Summarizing,  since our analysis is largely based on the uv/optical lines and continua, an A$_V$ intrinsic variability of about 1 mag might significantly reflect on the estimate of  physical parameters and mass accretion rate. Therefore, in the following we will consider conservatively the two extreme values of A$_V$, namely 1.5 and 2.9 mag. This will give us an idea of how much the A$_V$
variability influences the results.

\subsubsection{Mass accretion rate\label{sec:sec3.2.2}}

To determine the mass accretion rate (\macc) we applied two different methods (indeed not strictly independent) involving both  continuum and emission lines.

\begin{figure}
\epsscale{1.0}
\plotone{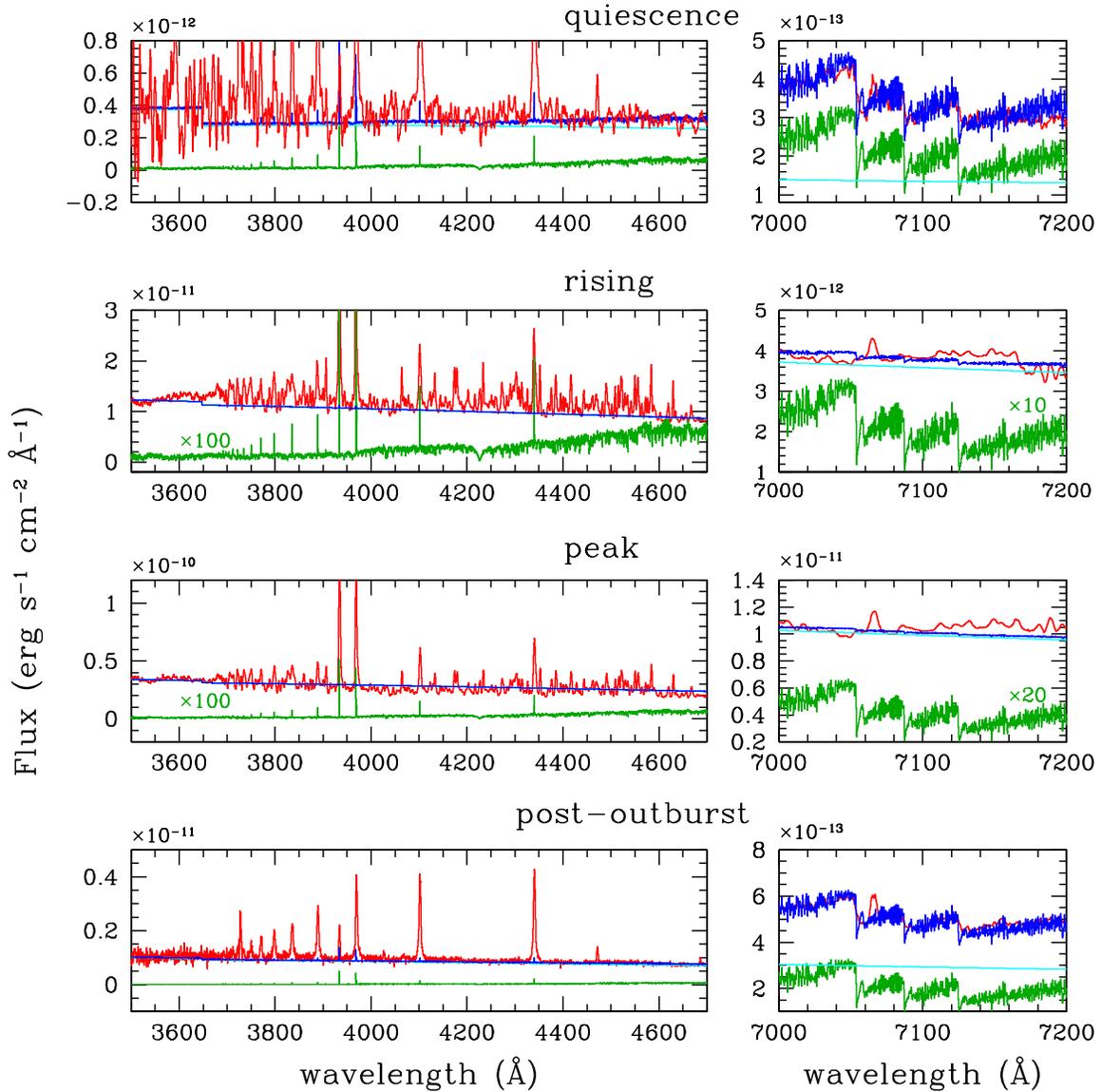}
\caption{Top panel: dereddened  spectrum (A$_V$\,=\,2.9 mag) of V1118 Ori between 3500$-$4700 \AA\, and 7000$-$7200 \AA, during quiescence (red). The spectrum of an M3 star template is shown in green. The continuum excess emission of the slab model is plotted in cyan, while the best fit with the emission predicted from the slab model added to the template is given by the blue line. Other panels: the same as in the top panel for the rising, peak, and post-outburst phases. In this latter, the spectrum was dereddened for A$_V$\,=\,1.9 mag. For a better visualization, the template spectrum has been multiplied by a constant where indicated.\label{fig:fig8}}
\end{figure}

In the first method, described in Manara et al. (2013), we fitted the observed MODS spectrum in various continuum regions from $\sim$3200 \AA\, to $\sim$7500 \AA\, using the sum of a photospheric template and a slab model to reproduce the continuum excess emission due to accretion. We adopted the reddening law by Cardelli et al. (1989) with $R_V$ = 3.1. We first fitted the quiescence spectrum taking as free both the stellar parameters and the extinction value. The best fit (Fig.~\ref{fig:fig8}, top panel), is obtained with a photospheric template of spectral type (SpT) M3 and for $A_V$ = 2.9 mag (as anticipated in the previous Section). These lead to \rstar\,=\,1.67 \rsun\, and $L_\star$ = 0.32 $L_\odot$, when a distance of 414 pc is adopted. By assuming the evolutionary models of Baraffe et al. (2015), the fitted parameters correspond to  \mstar\,=\,0.29 \msun\,. According to our working plan, we also fitted the quiescence spectrum adopting  $A_V$ = 1.5 mag (not shown in the Figure). The fit gives SpT\,=\, M3, \rstar\,=\,1.07 \rsun\,, $L_\star$ = 0.18 $L_\odot$, and  \mstar\,=\,0.29 \msun\,.

Fits of the subsequent phases (rising, peak, and post-outburst, Fig.~\ref{fig:fig8}, second, third, and fourth panel) were made by assuming the stellar parameters fitted in quiescence, and keeping $A_V$ and SpT fixed. We varied the contribution of the slab to the emission until a best fit was found. This is the same procedure as adopted to fit DR Tau by Banzatti et al. (2014). 
The fit of the rising and peak spectra, in particular, is very uncertain because the continuum emission is dominated by the accretion shock. 
The \macc\, derived with this 'Continuum' (C) method are listed in Table\,\ref{tab:tab8}, first and second column.

As a second method, we applied the empirical relationships derived by Alcal\'a et al. (2016) between line and accretion luminosity (\lacc)\, for a sample of young active T\,Tauri stars in Lupus. We assumed the same parameter values as for the 
continuum fitting, which were also taken to recompute the mass accretion rate in quiescence and rising phases (given in Papers\,I and II). This allows us to  meaningfully compare the results. 
We applied the Alcal\'a et al. relationships to the brightest \hi\, lines (those detected with S/N $\ga$ 10), which are the Balmer lines in the MODS and Asiago spectra, and the Paschen and Br$\gamma$ lines in the LUCI2 spectra\footnote{We avoid using the H$\alpha$ line, as suggested by  Alcal\'a et al., because self-absorption features and multiple components, likely arising from gas that is not strictly related with the accretion columns, strongly affect the observed profile (see Sect.\ref{sec:sec3.3}).}. Finally, we converted  \lacc\, into \macc\,by  applying the relationship reported by Gullbring et al. (1998). The results are summarized in Table\,\ref{tab:tab8}, third and fourth column ('Line' (L)
method).

\begin{figure}
\epsscale{1.0}
\plotone{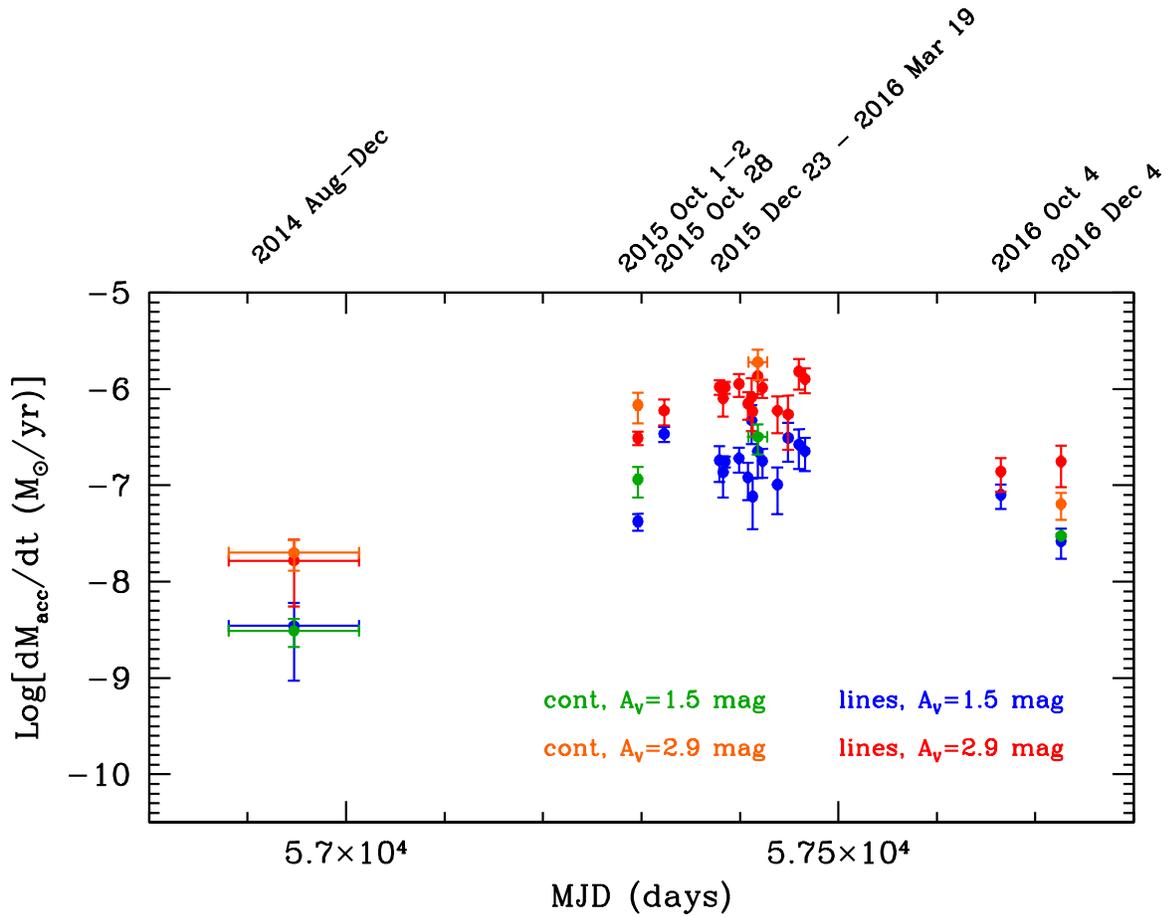}
\caption{Mass accretion rate as a function of time. With different colors we depict \macc\, values derived under different assumptions for the fitting method or the adopted extinction. 
Determinations of \macc\, that refer to 2014 August-December and 2015 October have been obtained from the spectra published in Papers\,I and 
II, respectively.\label{fig:fig9}}
\end{figure}

The \macc\, values are also shown in Figure\,\ref{fig:fig9}, where different colors are used to represent the adopted fitting method and extinction value. 
\macc\, increased by about two orders of magnitude from quiescence to the outburst peak, going from  0.3\,$-$\,2.0\,10$^{-8}$ \msunyr\ to 0.2\,$-$\,1.9\,10$^{-6}$ \msunyr\,. The upper end value is close to the estimate by Audard et al. (2010) at the peak of the 
2005 outburst, obtained by means of Spectral Energy Distribution (SED) fitting. Given the higher brightness attained by the 2005 outburst, its mass accretion rate should presumably be higher than for the 2015 event.

For a fixed A$_V$, the \macc\, derived with the two methods is consistent within the uncertainties, with the only exception that of the rising phase; conversely, differences of up to a factor of four are found between \macc\, derived under a different A$_V$ assumption. Since these differences are a due to  the {\it  intrinsic, short-timescale variability} of V1118 Ori, and not to a poor determination of A$_V$, the extreme \macc\, values have to be taken as the lower and upper end  values that 
characterize the considered temporal phase. 

Finally, it is also interesting to note that \macc\,(post-outburst) $\approx$ \macc\,(rising), even if the continuum level is very close to the quiescence value. This result is a direct consequence of the slower decrease of the 
line fluxes, as we have shown in Figure\,\ref{fig:fig2} for the case of the H$\beta$ flux.

\subsubsection{Hydrogen physical conditions\label{sec:sec3.2.3}}	 
\begin{figure}
\epsscale{1}
\plotone{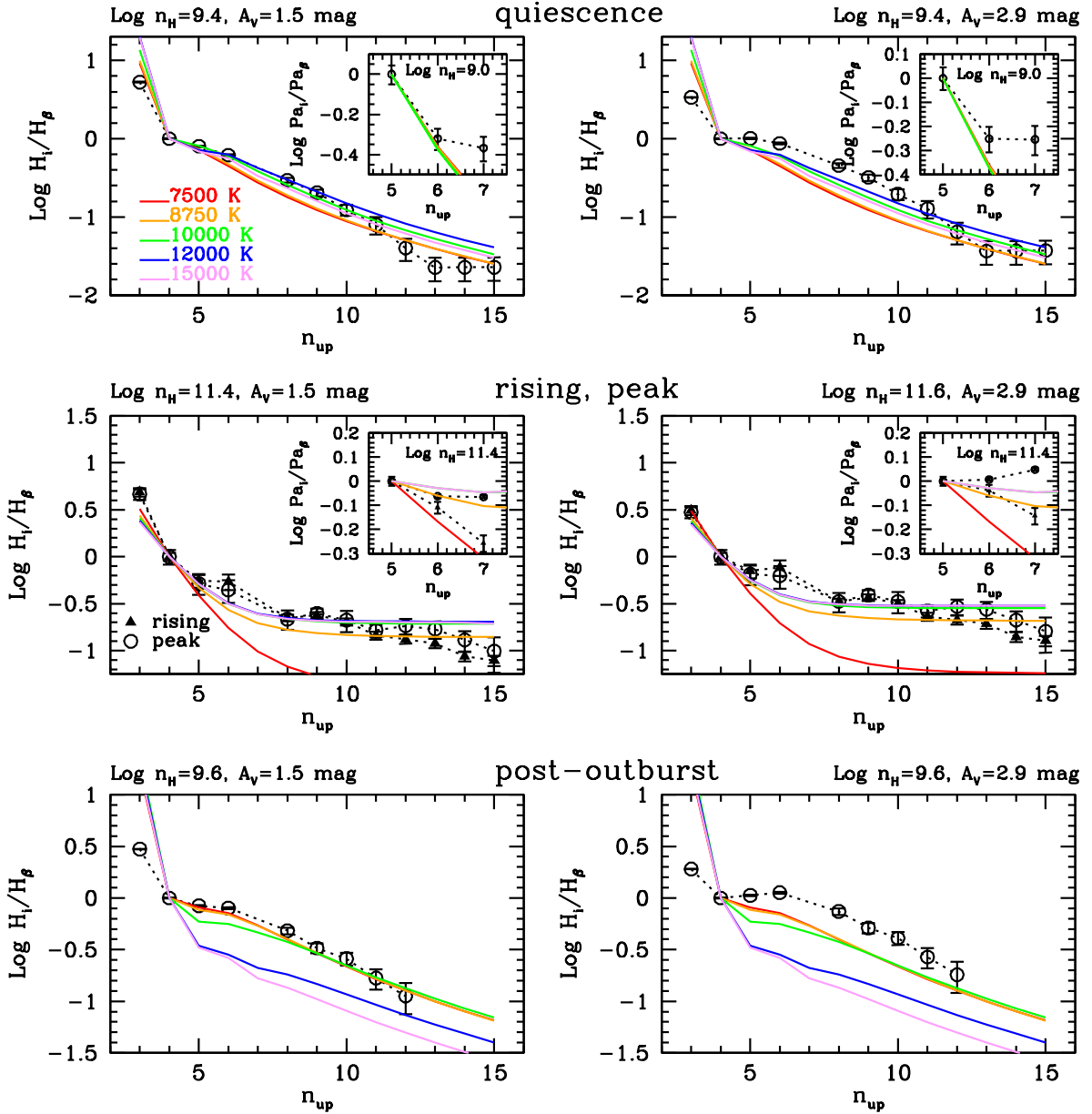}
\caption{Balmer decrements (H$_i$/H$\beta$ flux ratios vs. upper level of the transition) in quiescence (top panels), rising/peak (central panels), and post-outburst phases (bottom panels). In the left and right panels the observed ratios are corrected for A$_V$\,=\, 1.5 mag and 2.9 mag, respectively. The fluxes have been normalized to the H$\beta$ flux. The observations have been fitted with the local line excitation models of Kwan \& Fischer (2011). Models with different temperatures are depicted with different colors, and the adopted density is labeled. In the insets, the Paschen decrements (normalized to the Pa$\beta$) are shown and compared with the relative predictions of Edwards et al. (2013). \label{fig:fig10}}
\end{figure}

To derive the physical conditions of the gas that emits the \hi\, lines, we have compared the observed  Balmer  decrements (plots of flux ratios vs. upper level of the transition) with the predictions of the local line excitation computation by Kwan \& Fischer (2011), who provide a grid of models in
 the range of hydrogen density log [$n_{\rm{H}}$ (cm$^{-3}$)] = 8-12 and temperature between  3750 K and 15\,000 K. We have also examined the Paschen decrements for the few lines we observe, and compared them with the models by Edwards et al. (2013).
 We show the results in Figure\,\ref{fig:fig10}, where Balmer (from MODS spectra) and Paschen (from LUCI2 spectra) decrements are represented in the large panels and the insets, respectively. Since the H$\alpha$ line can be severely contaminated by opacity effects and chromospheric emission, we preferred to use the H$\beta$ as a reference line for the  Balmer decrements (e.g. Antoniucci et al. 2017). Furthermore, as the decrement shape may vary depending on the adopted reference line,
we have verified that using lines with a higher n$_{up}$ as reference does not provide significantly different results from those reported here.
The Paschen decrements have been normalized to the Pa$\beta$ line.

Balmer decrements during the quiescence phase are shown in the top panels, after they have been corrected for A$_V$\,=\,1.5 mag (left) and A$_V$\,=\,2.9 mag (right). Regardless of the 
adopted extinction value, the decrement shape resembles the {\it straight} type 2 shape cataloged by Antoniucci et al. (2017).  To this class of decrements typically belong T\,Tauri sources with  \macc\, in the range -10.8 $<$ log [\macc (\msunyr)]$<$ -8.8, in good agreement with our estimates of Section\,\ref{sec:sec3.2.2}. For the two A$_V$ values, we find that models with log [$n_{\rm{H}}$ (cm$^{-3}$)] = 9.4  are those that best fit the observed decrements, even if the overall shape is not well reproduced. No significant constraint can be placed on the hydrogen temperature from the Balmer decrements alone. While lower temperatures seem favored when A$_V$\,=\,1.5 mag is adopted, the opposite occurs for A$_V$\,=\,2.9 mag. The Paschen decrements, although not well fitted by models\footnote{Note that we consider Paschen models at log [$n_{\rm{H}}$ (cm$^{-3}$)] = 9.0  and log [$n_{\rm{H}}$ (cm$^{-3}$)] = 11.4 for quiescence and rising/peak, since models at the same density as the Balmer ones are not available in the literature.}, suggest $T \la$ 10\,000 K.

In the central panels we show the Balmer decrements of the rising (triangles) and peak (circles) phases. The most important difference with respect to quiescence is the enhancement of the ratios of lines with n$_{up}$ $\ga$ 10 of at least an order of magnitude. When compared with the decrement classification of Antoniucci et al., rising and peak decrements resemble the $L-shape$ type 4
decrements, observed mostly in objects with log [\macc (\msunyr)] $>$ -8.8. Again according to the Kwan \& Fischer models, the hydrogen density has increased by about two orders of magnitude with respect to quiescence, while temperatures between 8750 and 15\,000 K are suggested by both the Balmer and Paschen decrements. 

In the bottom panels we show the Balmer decrements of the post-outburst phase. Their shape is of  type 3, or ({\it bumpy}), according to the definition of Antoniucci et al., which is intermediate between types 2 and 4  (-9.8 $<$ log [\macc (\msunyr)]$<$ -8.8). A good agreement between models and data is found only for log [$n_{\rm{H}}$ (cm$^{-3}$)] = 9.6, T$\la$ 10\,000 K and A$_V$\,=\,1.5 mag.

We can conclude that the physical conditions of the gas profoundly changed  from quiescence to outburst, and this is especially true for the gas density. We can argue that this has led also to a similar enhancement in the accretion column density and consequently in the mass accretion rate.

\subsubsection{Other lines\label{sec:sec3.2.4}}

The high-resolution Asiago (echelle) and TNG (GIANO) spectra display a number of unblended emission lines of neutral atoms typical of M-type stars (e.g. Na, Ca, Fe, Ti, and Si, see Appendices A and B). Because of their low-ionization potential (between 5 and 8 eV), these lines likely originate in neutral regions 
close to the disk surface, far enough from the central star and accretion columns to prevent their ionization, but still 
reached by stellar UV photons able to populate energy levels close to the continuum, however. 
In Table\,\ref{tab:tab9} we list for each metallic species the energy of the most highly excited line, max(E$_{up}$), and the number of detected lines, N$_{lines}$. The reported statistics is done considering only the optical lines (in the 3300$-$7300 \AA\, range covered at high resolution), since this allows a meaningful comparison with the  
optical spectrum of EX Lup, taken during its 2008 outburst by Sicilia-Aguilar et al. (2012). As we show  in the last two columns of Table\,\ref{tab:tab9}, the EX Lup spectrum is definitely richer of high-excitation lines. This suggests that in this source the accretion shock produces more energetic photons or that the  radiation shielding is less efficient in its environment. It is interesting, however, that in contrast to this general trend, much more highly excited lines of \ion{Ti}{1}  and \ion{V}{1} are detected in V1118 Ori. Very likely, both these species are almost completely ionized in EX Lup, given their low-ionization potential (6.8 eV for \ion{Ti}{1} and 6.7 eV for \ion{V}{1}). Finally, we report the detection of \li\,\,6707 \AA\, in emission (S/N $\sim$ 4), a feature that is also present in the outburst of 2005 (Herbig 2008). This line is usually absent or seen in absorption in the spectra of classical T\,Tauri stars. 

\begin{figure}
\epsscale{1.0}
\plotone{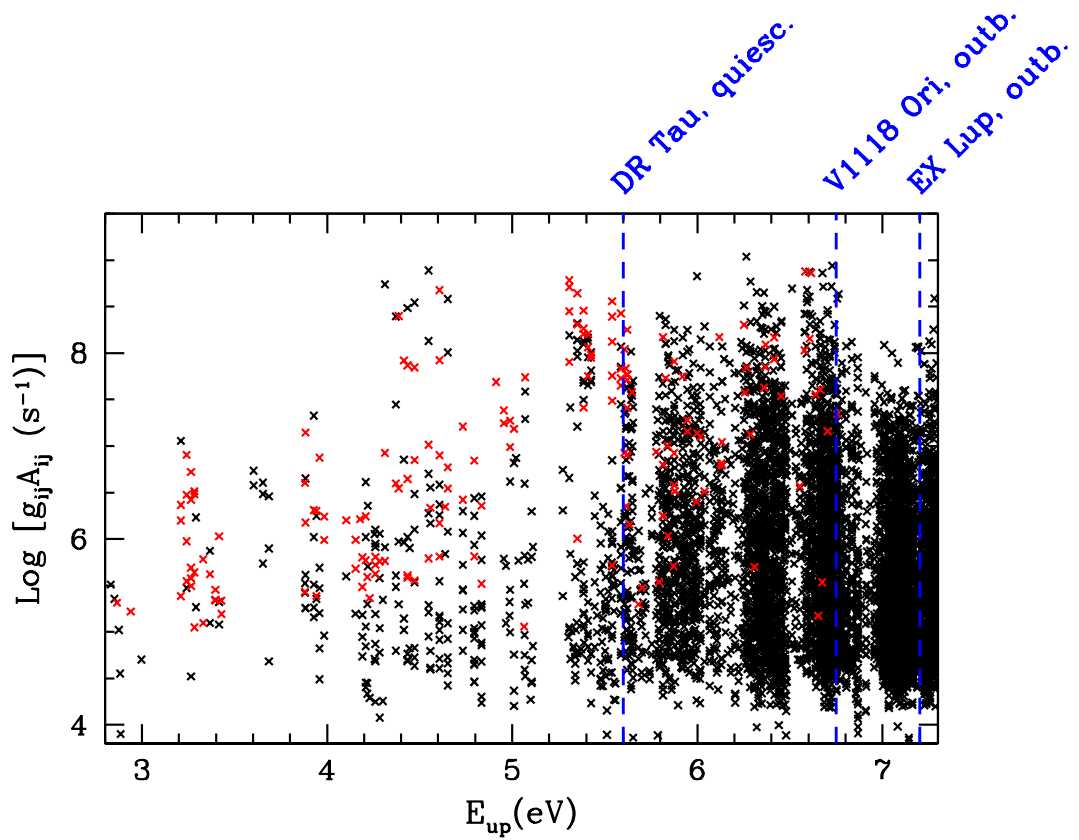}
\caption{Einstein coefficients $\times$ statistical weight (g$_{ij}$A$_{ji}$) of \ion{Fe}{1} lines between 4\,000 and 20\,000 \AA\, as a function of the energy of the upper level (black dots). Lines observed in V1118 Ori are represented in red. Blue lines indicate the energy of the line with the highest value of E$_{up}$ observed in DR Tau (Beristain et al. 1998), V1118 Ori, and EX Lup (Sicilia-Aguilar et al. 2012).\label{fig:fig11}}
\end{figure}

It is interesting to perform a more in-depth study of the  \ion{Fe}{1} lines that dominate the emission spectrum during the outburst. For this study we have considered {\it all} the \ion{Fe}{1} lines
predicted by theory with $\lambda$ between 4000 and 20\,000 \AA\,\,and inside the atmospheric windows, but discarding those with $A$ Einstein coefficients $\le$ 10$^3$ s$^{-1}$ or energy of the upper level E$_{up}$ $\ge$ 7.15 eV. The atomic data ($A-$coefficients, statistical weights and E$_{up}$ values) are taken from 'The atomic line list', v.2.05b21, http://www.pa.uky.edu/~peter/newpage/.
In Figure\,\ref{fig:fig11} we plot as black crosses the theoretical product g$_{ij}$A$_{ij}$ (statistical weight $\times$ Einstein coefficient) of the \ion{Fe}{1} lines  as a function of E$_{up}$. This product is proportional to the line intensity, and therefore it gives an idea of  the brightest lines depending on the level of the gas excitation. 
With red crosses we indicate the lines observed in V1118 Ori. Although with exceptions, we observe the lines having g$_{ij}$A$_{ij}$ $\ga$ 5 10$^5$ s$^{-1}$ for E$_{up}$ $\la$ 6.5 eV and g$_{ij}$A$_{ij}$ $\ga$ 10$^7$ s$^{-1}$ for E$_{up}$ $\ga$ 6.5 eV. We do not observe lines having E$_{up}$ $>$ 6.7 eV, 
a value in between the maximum  E$_{up}$ of lines observed in DR Tau  (in quiescence, Beristain et al 1998) and in EX Lup (during outburst, Sicilia-Aguilar et al. 2012).

Indications of the temperature and density of the line-emitting regions can be derived from the ratios of bright lines. The three \caii\, emission lines at $\lambda\lambda$ 8498, 8542, and 8662 \AA\, remained roughly constant throughout the outburst duration with ratios close to unity, which are suggestive of  optically thick environments. According to the analysis of a sample of Herbig stars, Hamann \& Persson (1992) have estimated that \caii\, optically thick lines are produced for an electron density of 10$^{11}$ cm$^{-3}$ and a hydrogen column density $\ge$ 10$^{21}$ cm$^{-2}$, if the temperature is between 4000 and 10\,000 K.  
Indeed, we can set as upper limit the maximum hydrogen temperature of $\sim$ 15\,000 K derived from the analysis of the Balmer lines (Sect.\ref{sec:sec3.2.3}), and as a lower limit the few thousand Kelvin (typically $\la$ 4000 K) that are typically 
traced by the emission of the CO ro-vibrational bands (see below).

Other remarkable lines are the \oxy\, triplet at 1.29 $\mu$m, pumped by the Ly$\beta$ (McGregor et al. 1988), and whose de-excitation originates the 8446\,\AA\,, and the \nai\, 2.206, 2.207\,$\mu$m doublet. Lorenzetti et al. (2009, their Figure 14) noted that in EXor spectra the \nai\, doublet is detected both in emission and in absorption following  the same behavior of the CO bands. Usually, 
strong CO emission is seen during outburst phases, while a progressive fading, absence, or absorption is detected as the outburst proceeds toward the quiescence (e.g. Aspin 2008 and references therein;  Biscaya et al. 1997;  Hillenbrand et al. 2013). 
This is also seen in V1118 Ori:  neither quiescent nor declining spectra show either CO bands or the \nai\,doublet, while in outburst both species are seen in emission  (Figure\,\ref{fig:fig5}). 
The common origin of CO and \nai\, emission is likely due to the very low-ionization potential of neutral sodium (5.1 eV), which can survive only close to low-mass late-type stars (like EXor, T\,Tauri, e.g. Antoniucci et al. 2008), and more generally, in regions deep enough inside the circumstellar disks where CO ro-vibrational bands are collisionally excited and 
shielding against photons that are able to produce Na$^{+}$ is efficient (McGregor et al. 1988). These conditions require 
densities $>$10$^7$ cm$^{-3}$  and only moderately warm temperatures, because CO is completely  dissociated at $\sim$4000 K.

The \oi\,6300\,\AA\, line is the only forbidden optical line present in the V1118 Ori spectrum (at an S/N $\sim$ 6$-$8).
In the Asiago high-resolution spectra, the line presents a structured profile that can be deconvolved  into two components, one close to the  peak of the diffuse \oi\,6300\,\AA\,  in the Orion nebula (v$_{helio} \sim$ +30 km s$^{-1}$, Garc{\'{\i}}a-D{\'{\i}}az \& Henney 2007), and a blue-shifted component centered at v$_{helio} \sim$ $-$15 km s$^{-1}$, likely originating in an outflowing wind. 
The line was also detected during the rising phase (Paper\,I), then increased by about a factor 4$-$5 in the peak spectra, and progressively faded in the following months. Hence, although the outflowing activity is not strictly connected with the accretion variability, it varies on the same timescales. 

Finally, we signal the detection of a faint H$_2$ v=1-0, 2.12\,$\mu$m line in the LUCI2 spectrum taken during the declining phase on 2016 October 4. It was most likely confused inside the continuum fluctuation in the peak spectra, but it was barely detected during quiescence and rising phases (Papers\,I and II).
Moreover, the line is resolved in the high-resolution GIANO spectra, where  it is  detected at an S/N $\sim$ 10. The heliocentric line peak is located at $\sim$ +36 km s$^{-1}$, and the FHWM is $\sim$ 32 km s$^{-1}$. Both these values are  close to the peak and FWHM of the diffuse \oi\,6300 \AA.\, This circumstance, added to the fact that the 2.12\,$\mu$m flux does not increase with the continuum (since the line was undetected in the peak spectra),
favors the hypothesis that the H$_2$ emission is related  to the diffuse cloud rather than to the local emission close to  V1118 Ori.

\subsection{Spectral profiles\label{sec:sec3.3}}

The kinematical evolution of the emitting gas during outburst was studied through the analysis of selected spectral profiles (those detected with the highest S/N). These have been deconvolved in a number of Gaussians using the curve-fitting tool PAN within the package DAVE\footnote{https://www.ncnr.nist.gov/dave/} (Azuah et al. 2009). Given the complexity of some profiles, for most of the lines the deconvolution  is not unique, since several fits with  comparable  $\chi^2$ can be found by varying the number of Gaussians and/or their parameters. The most complex profiles are exhibited by the H$\alpha$ and H$\beta$ lines. Their components change in intensity with time but remain substantially constant in number. The same is true for the other \hi\, lines, whose profiles can be accounted for with three or fewer Gaussians.

Given the evidence described above, we applied the following fitting procedure: (1) first, every line profile observed in the first date was fitted with the smallest possible number of Gaussians, letting free all the parameters (center, FWHM, area); (2) for each line, the output relative to the first date was taken as a guess solution for fitting the profile of the second date, maintaining fixed the number of Gaussians and allowing the parameters to vary within 30\%; (3) the same procedure was applied for all the dates of observations, considering as a guess solution that fitted in the previous date.  As a check of the validity of the adopted procedure, we also verified that the results did not change when lines with similar profiles, as for example H$\alpha$ and H$\beta$, were fitted simultaneously at a fixed date.

\begin{figure}
\epsscale{1.0}
\plotone{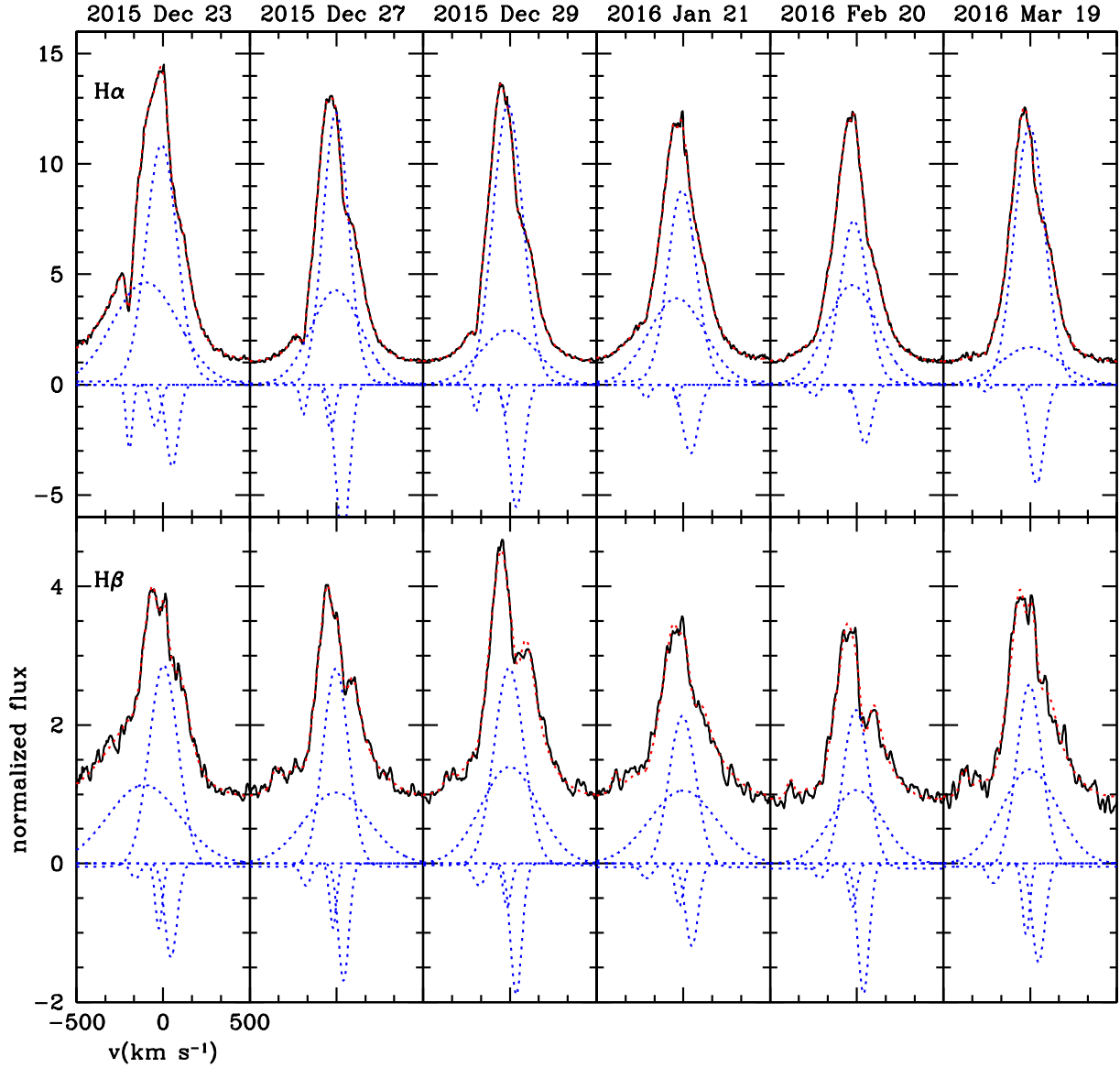}
\caption{H$\alpha$ and H$\beta$ continuum-normalized profile (black). In red we show the fit to the profile, obtained by adding multiple Gaussians (blue dotted lines, where a constant c=1 was subtracted from the fit for  better visualization). The date of the observation is reported as well.\label{fig:fig12}}
\end{figure}

\begin{figure}
\epsscale{0.8}
\plotone{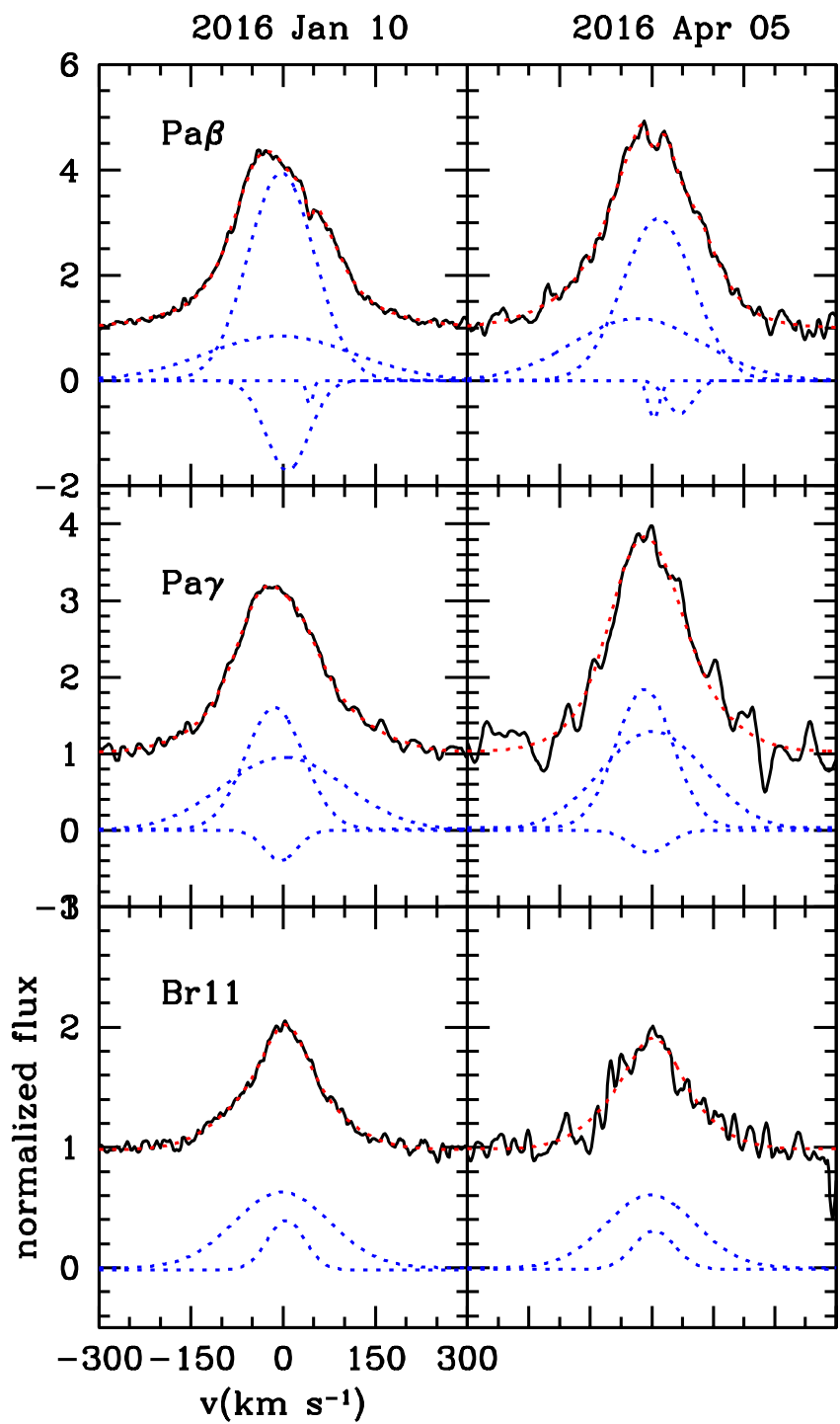}
\caption{Continuum-normalized profile of the brightest Paschen and Brackett lines (black). In red we show the fit to the profile, obtained by adding multiple Gaussians (blue dotted lines, where a constant c=1 was subtracted fromthe fit for better visualization). The date of the observation is reported as well. \label{fig:fig13}}
\end{figure}

Fits of selected lines are shown in Figures\,\ref{fig:fig12}, \ref{fig:fig13} and in Table\,\ref{tab:tab10}. Although
five Gaussians are required to fit the H$\alpha$ and H$\beta$ profiles, most of the observed flux is accounted for by two main components. These are a narrow component, slightly blue-shifted of a few km s$^{-1}$ with a FWHM close to 200 km s$^{-1}$, and a broad component, centered close to the rest velocity, with extended wings wider than $\pm$\,500 km s$^{-1}$. Variable self-absorption features are recognizable on both sides of the profile, and the red-shifted feature (centered at $\sim$ +40 km s$^{-1}$) is the most prominent. An outflowing wind is shown by a  P Cyg-like absorption (e.g. Edwards et al. 2003) at v $\sim$ $-$200 km s$^{-1}$, clearly visible in the H$\alpha$ profile on 2015 December 23. In contrast to the observations of Herbig (2008), who noted a link between the source activity and  a P Cyg absorption, our spectra reveal a more complex situation, since the P Cyg clearly faded in a few days while the outburst was in a stable and peak phase. Similarly, the H$\beta$, although not presenting a prominent P Cyg-like feature, exhibits
a pronounced blue wing in the spectrum of December 23 that disappears in the following days. Such sudden and brief wind enhancement is also shown by the shift of the broad velocity component peak of about $-$100  km s$^{-1}$ (see Table\,\ref{tab:tab10}). 

\begin{figure}
\epsscale{1.0}
\plotone{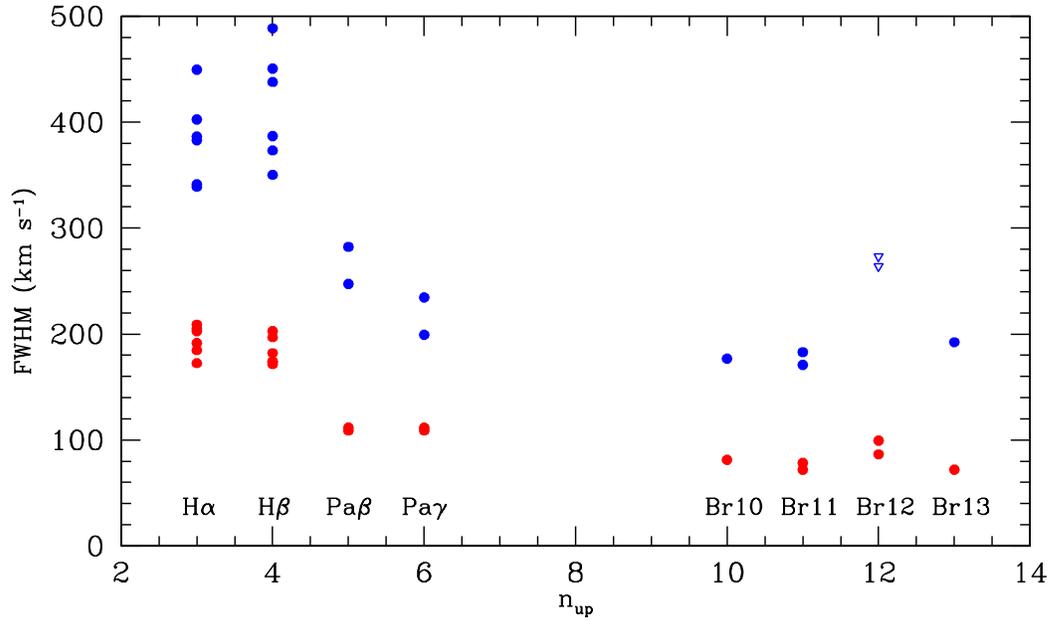}
\caption{FHWM of the brigthest \hi\, lines plotted against n$_{up}$. Red and blue dots represent the narrow and broad component, respectively. Triangles indicate upper limits on the FWHM of the Br12 line (see text). Multiple points of the same line refer to different dates of observation.\label{fig:fig14}}
\end{figure}

In contrast to the Balmer lines, other \hi\, lines present more symmetric profiles (Figure\,\ref{fig:fig13}) composed of a narrow and a broad component,
 which tend to become narrower with the increase of the number of the upper energy level (n$_ {up}$, Figure\,\ref{fig:fig14})\footnote{The only exception is
  the profile of the Br12 line, whose broad component is contaminated by Ti\,I and V\,I lines.}. First, this might be explained in the framework of a
   temperature and density gradient in the emitting gas. For example, according to the Kwan and Fischer models of the Balmer decrements (Figure\, \ref{fig:fig10}), flux ratios involving lines with higher n$_{up}$ increase with increasing temperature and density. Since this hot and dense gas is probably only a portion of the total hydrogen
emitting gas, we expect that it is characterized by a smaller range of velocity than the gas probed by lines with lower n$_ {up}$ (e.g. H$\alpha$).
Other possible causes for differential line broadening may be Stark and van der Waals effects, which may produce significant high-velocity emission in H$\alpha$ and, to a lesser extent, in other Balmer lines (Muzerolle, Calvet, \& Hartmann  2001).
We cannot exclude, however, that the observed behavior is, at least partially, a consequence of other causes. In particular,  the FHWM of the narrow component of the brightest lines might be overestimated as the profiles can be saturated or even self-absorbed because of opacity effects. Conversely, in the broad component, the FWHM of the faintest lines may be underestimated because the wings are confused inside the continuum fluctuation.

\begin{figure}
\epsscale{0.8}
\plotone{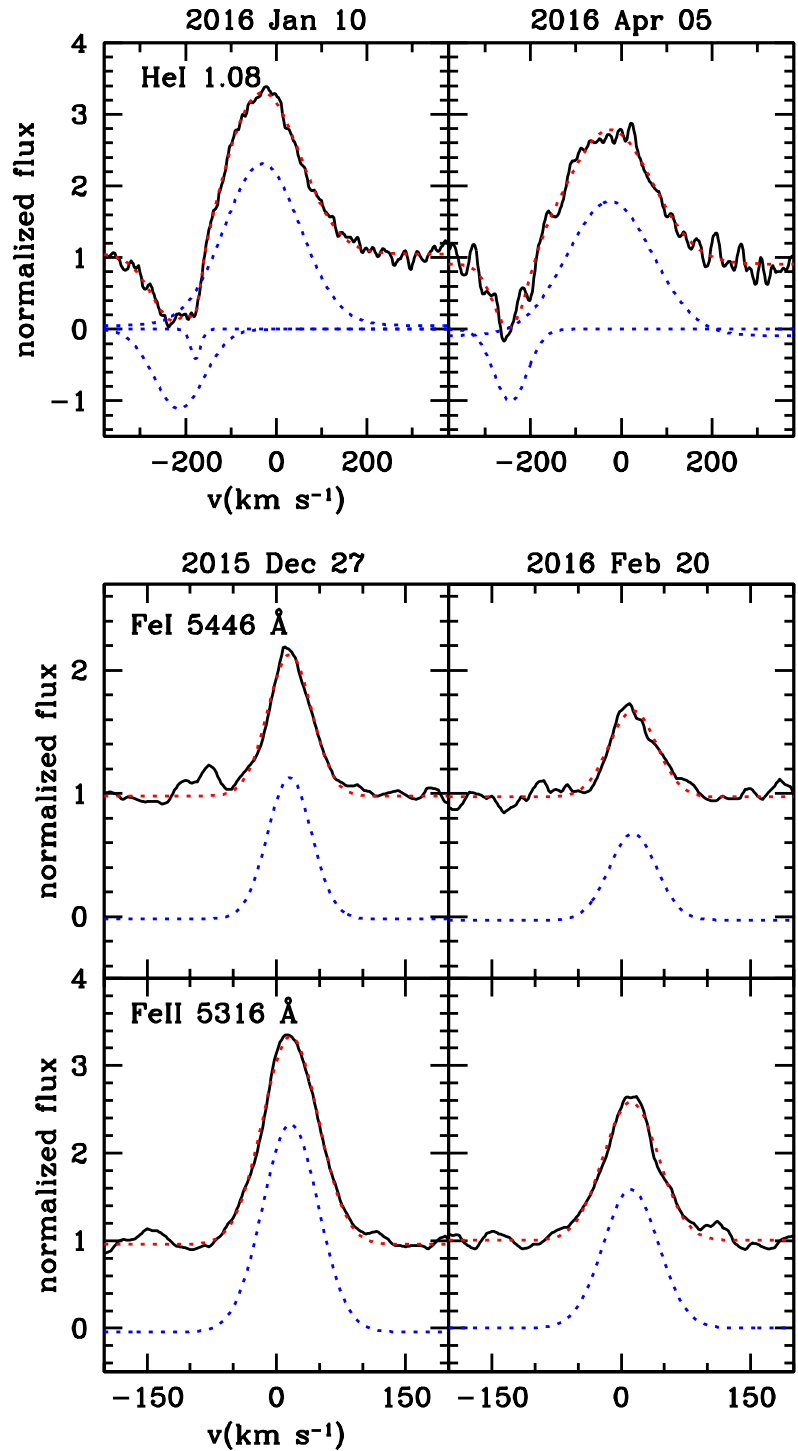}
\caption{Top panel: Continuum-normalized \hei\, 1.08\,$\mu$m profile (black). In red we show the fit to the profile, obtained by adding multiple Gaussians (blue dotted lines, where a constant c=1 was subtracted from the fit for a better visualization). The date of the observation is reported, as well. Bottom panels: the same as in the top panel for \fei\,\,5446 \AA\, and \feii\,\,5316 \AA\, taken as examples of the metallic lines. These profiles are shown on two dates that are representative of the outburst evolution. \label{fig:fig15}}
\end{figure}

In Figure\,\ref{fig:fig15}, top panel, we show the profile of the \hei\, 1.08 $\mu$m. The line can be deconvolved into  two main Gaussians in the two observation dates in 2016 January and April. The main 
component is in emission, with the center a v $\sim$ $-$25 km s$^{-1}$ and FHWM $\sim$ 200 km s$^{-1}$, namely similar to the parameters of the narrow component of the H$\alpha$ line. An ionized wind  is shown by a pronounced P Cyg-like absorption  (v$_{peak}$ $\sim$ $-$200 km s$^{-1}$ ), which, differently from the \hi\, lines, is well visible until the end of the peak phase.

\fei\, and \feii\, lines are well fitted with a single Gaussian centered close to the rest velocity with width $<$ 100 km s$^{-1}$ (Figure\,\ref{fig:fig15}, bottom panels). This
allows us to rule out both that they form in the stellar photosphere (the FHWM should be of only a few km s$^{-1}$) and that they originate from the same gas as the \hi\, and the \hei\, lines. Rather,
as already hypothesized in Section\,\ref{sec:sec3.2.4}, they might arise at the base of the accretion columns close to the disk surface, where the gas is mostly neutral and still relatively slow. 
Given the large number of \fei\, and \feii\, lines, we have investigated whether or not a difference exists between their average width ($<$FHWM$>$). We find that \fei\, lines  
have $<$FHWM$>$ = 64$\pm$8 km s$^{-1}$ against a $<$FHWM$>$ of \feii\, lines of 85$\pm$15 km s$^{-1}$.  For both species, the $<$FHWM$>$ evolves with time, and the maximum value is reached on 2016 January,
when the  $<$FHWM$>$ was 71$\pm$11 km s$^{-1}$ and 109$\pm$33 km s$^{-1}$ for the \fei\, and \feii\, lines, respectively.

\section{Summary\label{sec:sec4}}
We have presented a photometric and spectroscopic study of the recent outburst of the classical EXor source V1118 Ori, after following its evolution
 from the rising to the peak and declining phase. The derived parameters have been meaningfully compared with those already measured in quiescence at a similar level of sensitivity.
 Our main results  are summarized in Table\,\ref{tab:tab11}, and can be listed as follows:

\begin{itemize}
\item[1.] The 2015-2016 outburst of V1118 Ori is the sixth well-documented event in the past $\sim$ 40 years. 
The source  brightness suddenly increased  around 2015 September, reaching its peak around 2016 January and starting to decline around 2016 April. Magnitudes close to the quiescence values have been measured since 2016 September.
\item[2.] The 2015 outburst has occurred after a relatively long period of quiescence (about 10 years), while the events of the past three decades 
have shown a roughly regular recurrence of 6$-$8 years. This circumstance does not support the theory that periodic phenomena are the cause of the outburst triggering.
\item[3.] A blueing effect is  clearly recognizable in the [$V-R$] vs. [$R-I$] diagram. In agreement with previous works on the same object and on EXors in general, we interpret it as the  signature of the accretion burst.
\item[4.] Mass accretion rates were computed for all the outburst phases and compared with the quiescence value. We estimate that \macc\, increases by about two orders of magnitude from the quiescence to the outburst
 peak, specifically from 0.3$-$2.0 10$^{-8}$ \msunyr\, to 0.2-1.9 10$^{-6}$ \msunyr. A comparable increase occurs in the hydrogen density, from $\sim$ 2 10$^9$ cm$^{-3}$ to $\sim$ 4 10$^{11}$ cm$^{-3}$. The gas temperature is not well constrained, but we can place a lower limit of $\sim$ 8000\,$-$\,9000 K during outburst.  
\item[5.] Optical/near-infrared outburst spectra are rich in metallic lines, mainly coming from neutral atoms typical of M-type stars. Our analysis suggests that these lines originate close to the disk surface, in regions sufficiently shielded from UV radiation to prevent their ionization. This view is confirmed by the relatively 
narrow profiles ($\sim$ 60 km s$^{-1}$) of the \fei\, lines.
\item[6.] H$\alpha$ and H$\beta$ lines present complex profiles, with wings  extending up to $\pm$ 500 km s$^{-1}$, while profiles of other  \hi\, lines are more symmetric and tend to become narrower for increasing n$_{up}$. We tentatively explain such behavior considering that lines at higher energy originate preferentially in the warmest and densest gas component. Stark and van der Waals effects may also play a role. 
\item[7.]  An outflowing activity is demonstrated by a P Cyg-like 
absorption in the H$\alpha$, H$\beta$, and \hei\, 1.08\,$\mu$m profiles. While this feature disappears in the \hi\,line profiles on a timescale of a week, it is more stable in the highly ionized gas, being recognizable 
for at least four months in the \hei\, profile.\\ 

As a final consideration, we remark that our study has taken benefited greatly from the observations of the quiescence phase, since this
allowed us to define the {\it zeroes} of the physical parameters. Therefore systematic observations of EXors {\it also during quiescence} (like our EXORCISM monitoring program, which has been ongoing since several years), are of great help in order to place the outburst phenomenology in a comparative and statistical context.
\end{itemize}

\appendix
\section{Appendix A \label{sec:appendix}}
In Table \ref{tab:tab12} we give the list of the optical, unblended lines detected with the echelle at the 1.82 Asiago telescope ($\Re$ $\sim$ 20\,000).

\section{Appendix B \label{sec:appendix_b}}
In Table \ref{tab:tab13} we give the list of the near-infrared, unblended lines detected with GIANO at the TNG telescope ( $\Re$ $\sim$ 50\,000).

\section{Acknowledgements}
The authors sincerely thank John Kwan for providing the predictions of his local line excitation models. CFM acknowledges an ESA Research Fellowship.
This work is based on observations made with different instruments:[1] the Large Binocular Telescope (LBT). The LBT is an international collaboration among institutions in the United States, Italy and Germany. LBT Corporation partners are: The University of Arizona on behalf of the Arizona university system; Istituto Nazionale di Astrofisica, Italy; LBT Beteiligungsgesellschaft, Germany, representing the Max-Planck Society, the Astrophysical Institute Potsdam, and Heidelberg University; The Ohio State University, and The Research Corporation, on behalf of The University of Notre Dame, University of Minnesota and University of Virginia; [2] the Asiago ANS Collaboration telescopes operated under the responsability of the INAF-Osservatorio Astronomico di Padova (OAPd); [3] the AZT-24 IR telescope at Campo Imperatore (L'Aquila - Italy) operated by INAF-Osservatorio Astronomico di Roma (OAR); [4] the TNG telescope at La Palma (Spain) operated by INAF.
We acknowledge the observers that contribute to the AAVSO International Database.

{}

\

\begin{deluxetable}{ccccccc}
\tablewidth{0pt}
\tabletypesize{\footnotesize}
\tablecaption{\label{tab:tab1} Optical photometry} 
\tablehead{
\colhead{Date}        &  \colhead{MJD}      &   \colhead{B}        &   \colhead{V}      &   \colhead{R$_c$}        &  \colhead{I$_c$} & \colhead{Tel.}    
}
\startdata
                     & (days)    & \multicolumn{4}{c}{(mag)}                   &    \\
\cline{1-7}\\[-5pt]              
2016 Jan 06 & 57393.412 & 14.59     & 13.83	    & 13.24	 & 12.51      & 036\\
2016 Jan 13 & 57400.420 & 14.59     & 13.92	    & 13.28	 & 12.56      & 036\\
2016 Jan 14 & 57401.371 & 14.66     & 14.08	    & 13.39	 & 12.68      & 157\\
2016 Jan 16 & 57403.251 & 14.51     & 13.89	    & 13.18	 & 12.53      & 157\\
2016 Jan 17 & 57404.352 & 14.79     & 14.01	    & 13.38	 & 12.63      & 036\\
2016 Jan 20 & 57407.357 & 14.81     & 14.14	    & 13.51	 & 12.76      & 036\\ 
2016 Jan 26 & 57413.230 & 14.43     & 13.75	    & 13.09	 & 12.41      & 157\\ 
2016 Jan 26 & 57413.245 & 14.51     & 13.78	    & 13.18	 & 12.43      & 036\\ 
2016 Feb 01 & 57419.427 &  -        & 14.08     & 13.48	 & 12.74      & 036\\ 
2016 Feb 06 & 57424.234 & 14.96     & 14.28	    & 13.52	 & 12.83      & 157\\ 
2016 Feb 06 & 57424.253 & 14.89     & 14.26	    & 13.56	 & 12.82      & 036\\ 
2016 Feb 12 & 57430.236 & 14.89     & 14.32	    & 13.63	 & 12.95      & 157\\ 
2016 Feb 19 & 57437.242 & 14.97     & 14.49	    & 13.86	 & 13.09      & 157\\ 
2016 Feb 20 & 57438.243 & 14.84     & 14.17	    & 13.40	 & 12.74      & 157\\ 
2016 Feb 21 & 57439.296 & 14.53     & 13.72	    & 13.06	 & 12.47      & 036\\ 
2016 Feb 21 & 57439.312 & 14.26     & 13.74	    & 13.17	 & 12.49      & 157\\ 
2016 Feb 21 & 57439.362 & -         & 13.71	    &	-	 & 12.46      & 036\\ 
2016 Feb 22 & 57440.259 & 14.59     & 14.09	    & 13.20	 & 12.55      & 157\\ 
2016 Feb 22 & 57440.364 & 14.45     & 13.75	    & 12.97	 & 12.42      & 036\\ 
2016 Feb 24 & 57442.249 & 14.46     & 13.85	    & 13.14	 & 12.48      & 157\\ 
2016 Mar 02 & 57449.244 & 14.72     & 13.98	    & 13.36	 & 12.63      & 157\\ 
2016 Mar 02 & 57449.319 & 14.59     & 14.00	    & 13.38	 & 12.61      & 036\\ 
2016 Mar 15 & 57462.257 & 14.18     & 13.53	    & 12.84  & 12.17      & 157\\ 
2016 Mar 15 & 57462.281 & 14.23     & 13.58	    & 12.88	 & 12.19      & 036\\ 
2016 Mar 18 & 57465.256 & 14.28     & 13.56	    & 12.93	 & 12.26      & 157\\ 
2016 Mar 19 & 57466.258 & 14.38     & 13.72	    & 13.01	 & 12.29      & 157\\ 
2016 Mar 19 & 57466.316 & 14.43     & 13.67	    & 13.00	 & 12.30      & 036\\ 
2016 Mar 24 & 57471.301 & 14.55     & 13.79     &   -    & 12.41      & 036\\
2016 Mar 30 & 57477.284 & 14.42     & 13.75     & 13.08  & 12.39      & 157\\
2016 Apr 12 & 57490.279 & 14.58     & 13.80     & 13.06  & 12.36      & 157\\
2016 Apr 16 & 57494.285 & 14.65     & 13.84     & 13.14  & 12.42      & 157\\
2016 Aug 25 & 57625.622 & -         & 14.50     &  -     & 13.59      & 036\\
2016 Sep 06 & 57637.619 & 15.97     & 15.64     & 14.94  & 14.16      & 157\\
2016 Sep 09 & 57640.637 & 16.51     & 16.10     & 15.28  & 14.31      & 157\\
2016 Oct 27 & 57688.675 &  -        & 15.99     &  -     & 14.35      & 036\\
2016 Oct 29 & 57690.455 & 16.42     & 16.34     & 15.36  & 14.45      & 157\\
2016 Nov 11 & 57699.466 & 16.81     & 16.54     & 15.41  & 14.64      & 157\\
2016 Nov 30 & 57723.435 & 16.97     & 16.41     & 15.52  & 14.64      & 157\\
2016 Dec 05 & 57728.473 &  -        & 16.28     &   -    & 14.43      & 036\\
\cline{1-7}
\enddata
\tablecomments{Typical errors are smaller than 0.02 mag.}
\end{deluxetable}


\begin{deluxetable}{ccccc}
\tablewidth{0pt}
\tabletypesize{\footnotesize}
\tablecaption{\label{tab:tab2} Near-IR photometry.} 
\tablehead{
\colhead{Date}        &  \colhead{MJD}      &   \colhead{J}        &   \colhead{H}      &   \colhead{K}  
}
\startdata
         	    & (days)     &     \multicolumn{3}{c}{(mag)}  \\
\cline{1-5} 
2015 Dec 18     &57374.474	&    11.34    &  10.63   &   9.83 \\
2016 Feb 02     &57442.356	&    11.45    &  10.69   &   9.89 \\
2016 Feb 24     &57442.366	&    11.46    &  10.68   &    -   \\
2016 Sep 21     &57652.574	&    12.49    &  11.41   &  10.57 \\
2016 Sep 23     &57654.600	&    12.53    &  11.47   &  10.71 \\
2016 Sep 25     &57656.616	&    12.58    &  11.51   &  10.76 \\
2016 Sep 27     &57658.612	&    12.56    &  11.51   &  10.74 \\
2016 Sep 28     &57659.597	&    12.55    &  11.51   &  10.75 \\
2016 Sep 30     &57661.580	&    12.46    &  11.41   &  10.62 \\
2016 Oct 01     &57662.667	&    12.57    &  11.50   &  10.67 \\
2016 Oct 29     &57690.548  &    12.42    &  11.33   &  10.67 \\
2016 Oct 31     &57692.598  &    12.46    &  11.41   &  10.65 \\
2016 Nov 02     &57694.504  &    12.47    &  11.39   &  10.61 \\
\cline{1-5}
\enddata
\tablecomments{Typical errors are smaller than 0.03 mag.}
\end{deluxetable}


\begin{deluxetable}{ccccc}
\tablewidth{0pt}
\tabletypesize{\footnotesize}
\tablecaption{\label{tab:tab3} WISE photometry.} 
\tablehead{
\colhead{Date}        &  \colhead{MJD}     &   \colhead{[3.4]}          &   \colhead{[4.6]}         &   \colhead{[12]}       
}
\startdata
           &  (days)  & \multicolumn{3}{c}{(mag$\pm$ $\Delta$mag)}       \\
\cline{1-5}               
2010 Aug 03  & 55263   & 9.84$\pm$0.03    & 9.01$\pm$0.02	 & 6.87$\pm$0.06\\
2010 Sep 09  & 55454   & 9.95$\pm$0.05    & 9.10$\pm$0.06	 & 6.96$\pm$0.06\\
2014 Mar 08  & 56724   & 9.93$\pm$0.03    & 9.14$\pm$0.03	 &  -           \\
2014 Mar 11  & 56727   &10.04$\pm$0.17    & 9.21$\pm$0.09	 &  -           \\
2014 Mar 12  & 56728   & 9.94$\pm$0.02    & 9.12$\pm$0.03	 &  -           \\
2014 Sep 15  & 56915   & 9.95$\pm$0.03    & 9.13$\pm$0.03	 &  -           \\
2014 Sep 19  & 56919   & 9.96$\pm$0.03    & 9.12$\pm$0.03	 &  -           \\
2015 Mar 06  & 57087   & 9.94$\pm$0.03    & 9.13$\pm$0.02	 &  -           \\
2015 Sep 16  & 57281   & 8.98$\pm$0.06    & 8.14$\pm$0.06	 &  -           \\
\cline{1-5}
\enddata
\end{deluxetable}

\begin{deluxetable}{ccccccc}
\tablewidth{0pt}
\tabletypesize{\footnotesize}
\tablecaption{\label{tab:tab4} Journal of spectroscopic observations.}
\tablehead{
\colhead{Date}	& \colhead{MJD} & \colhead{Telescope/Instrument}&  \colhead{$\Delta\lambda$} & \colhead{$\Re$}   &  \colhead{rms} &\colhead{ Exp time}  
}
\startdata
         	    &             &                         &  (\AA)          &              & ($10^{-16}$ erg s$^{-1}$  cm$^{-2}$ \AA$^{-1}$)   &  (s)      \\
\cline{1-7}   
2015 Dec 23     & 57379       & 1.22m/B\&C               &   3300-8050     & 2400         & 7-12                                      & 1800       \\ 
2015 Dec 23     & 57379       & 1.82m/ECH                &   3600-7300     & 20\,000      & 7-20                                      & 1800       \\
2015 Dec 27     & 57383       & 1.22m/B\&C               &   3300-8050     & 2400         & 4-7                                       & 3600       \\
2015 Dec 27     & 57383       & 1.82m/ECH                &   3600-7300     & 20\,000      & 7-25                                      & 1800       \\
2015 Dec 29     & 57385       & 1.82m/ECH                &   3600-7300     & 20\,000      & 4-13                                      & 3600       \\
2016 Jan 10     & 57397       & TNG/GIANO                &   9500-24\,500  & 50\,000      & -                                         & 3000       \\
2016 Jan 12     & 57399       & LBT/MODS                 &   3200-9500     & 1500         & 0.9-1.2                                   & 1500       \\    
2016 Jan 21     & 57408       & 1.82m/ECH                &   3600-7300     & 20\,000      & 15-38                                     & 3600       \\
2016 Jan 21     & 57408       & 1.22m/B\&C               &   3300-8050     & 2400         & 3-6                                       & 3600       \\
2016 Jan 25     & 57412       & LBT/LUCI2                &  10\,000-24\,000& 1000         & 0.7-1.6                                   & 1200       \\
2016 Jan 26     & 57413       & LBT/MODS                 &   3200-9500     & 1500         & 0.9-1.1                                   & 1500       \\      
2016 Jan 31     & 57418       & LBT/MODS                 &   3200-9500     & 1500         & 0.7-1.0                                   & 1500       \\      
2016 Feb 05     & 57423       & 1.22m/B\&C               &   3300-8050     & 2400         & 4-7                                       & 3600       \\
2016 Feb 20     & 57438       & 1.22m/B\&C               &   3300-8050     & 2400         & 2-6                                       & 5400       \\
2016 Feb 20     & 57438       & 1.82m/ECH                &   3600-7300     & 20\,000      & 12-41                                     & 3600       \\    
2016 Mar 02     & 57449       & LBT/LUCI2                &  10\,000-24\,000& 1000         & 1.1-2.2                                   & 1200       \\
2016 Mar 13     & 57460       & 1.22m/B\&C               &   3300-8050     & 2400         & 10-20                                     & 900        \\
2016 Mar 19     & 57466       & 1.82m/ECH                &   3600-7300     & 20\,000      & 12-42                                     & 1800       \\
2016 Mar 19     & 57466       & 1.22m/B\&C               &   3300-8050     & 2400         & 8-20                                      & 2400       \\
2016 Apr 05     & 57483       & TNG/GIANO                &   9500-24\,500  & 50\,000      & -                                         & 3000       \\
2016 Oct 04     & 57665       & LBT/LUCI2                &  10\,000-24\,000& 1300         & 0.7-2.0                                   & 1200       \\
2016 Dec 04     & 57726       & LBT/MODS                 &   3250-9500     & 1500         & 0.7-1.0                                   & 1500       \\  
\cline{1-7}
\enddata
\end{deluxetable}

\begin{deluxetable}{cccccc}
\tablewidth{0pt}
\tabletypesize{\footnotesize}
\tablecaption{\label{tab:tab5} Fluxes of main lines detected with LBT/MODS.} 
\tablehead{
\colhead{ } &    \colhead{ }            &  \colhead{2016 Jan 12}     &   \colhead{2016 Jan 26}       &    \colhead{2016 Jan 31}     &    \colhead{2016 Dec 04}    
}
\startdata           
Line        & $\lambda_{air}$   &   \multicolumn{4}{c}{Flux$\pm \Delta$ Flux}                                     \\
            &  (\AA)            &   \multicolumn{4}{c}{ (10$^{-15}$ erg s$^{-1}$ cm$^{-2}$)}                      \\
\cline{1-6}              
  H15    	&  3711.97          & 10.5$\pm$1.8     &    -	              &    15.8$\pm$2.0     &   -             \\
  H14    	&  3721.94	        & 14.3$\pm$1.8     &   9.9$\pm$0.9		  &    18.8$\pm$2.0     &   -             \\
  H13    	&  3734.37          & 18.7$\pm$1.8     &  12.2$\pm$1.1	  	  &    23.9$\pm$2.0     &   -             \\
  H12    	&  3750.15     	    & 21.7$\pm$1.8     &  12.7$\pm$1.1	      &    25.5$\pm$2.0     & 2.1$\pm$0.7     \\
  H11	    &  3770.63      	& 20.4$\pm$1.8     &  11.6$\pm$1.1	      &    22.4$\pm$2.0     & 3.1$\pm$0.7     \\
  H10	    &  3797.90      	& 21.9$\pm$1.8     &  10.9$\pm$1.1	      &    31.6$\pm$2.0     & 4.9$\pm$0.7     \\
  H9	    &  3835.38      	& 32.8$\pm$1.8     &  10.7$\pm$1.1	      &    32.5$\pm$2.0     & 6.3$\pm$0.7     \\
  H8        &  3889.05          & 29.6$\pm$1.8     &  10.3$\pm$0.9        &    33.0$\pm$2.0     & 9.5$\pm$0.7     \\
 CaII H     &  3933.66          & 105.92$\pm$1.7   &  73.3$\pm$1.0        &   170.3$\pm$2.0     & 3.5$\pm$0.7     \\
 CaII K + H7  &  3968.45/3970.07  & 99.2$\pm$1.7   &  55.3$\pm$1.0        &   152.4$\pm$2.0     & 14.1$\pm$0.7    \\
 HeI        &  4026.19          &  9.2$\pm$1.7     &   6.5$\pm$1.0        &     9.8$\pm$2.0     & -               \\
 H$\delta$  &  4101.73          & 50.4$\pm$1.7     &  28.3$\pm$1.0        &    75.7$\pm$2.0     & 17.15$\pm$0.4   \\
 H$\gamma$  &  4340.46          & 71.9$\pm$1.7     &  42.0$\pm$1.0        &    96.8$\pm$2.0     & 20.3$\pm$0.4    \\
 H$\beta$   &  4861.32          & 186.11$\pm$1.7   & 112.6$\pm$1.0        &   231.3$\pm$2.0     & 31.1$\pm$0.2    \\
 HeI        &  5015.68          & 16.18$\pm$1.7    &  12.0$\pm$1.0        &    14.1$\pm$2.0     & 1.4$\pm$0.2     \\
 Mg I       &  5168.76/5174.12  & 27.8$\pm$1.7     & 36.2$\pm$1.7         &  20.2$\pm$1.0       &   -             \\
 Mg I       &  5185.05          & 7.3$\pm$1.7      & 10.2$\pm$1.7         &    5.4$\pm$1.0      &   -             \\
 HeI        &  5875.6           & 27.1$\pm$1.8     &  13.5$\pm$1.1        &    29.5$\pm$2.0     & 3.5$\pm$0.2     \\
 $[\rm{OI}]$&  6300.30          & 15.5$\pm$1.8     &   9.5$\pm$1.2        &    11.7$\pm$2.0     & 15.75$\pm$0.2   \\
 H$\alpha$  &  6562.80          & 1546.8$\pm$2.0   & 765.0$\pm$1.2        &  1635.5$\pm$2.0     & 149.4$\pm$0.2   \\
 HeI        &  6678.15          & 23.9$\pm$2.0     &  14.5$\pm$1.2        &    23.2$\pm$2.0     & 2.8$\pm$0.2     \\
 HeI        &  7065.2           & 15.8$\pm$2.0     &  10.3$\pm$1.2        &    14.3$\pm$2.0     &   -             \\
 OI         &  8446.5           & 59.6$\pm$2.0     &  34.4$\pm$1.1        &    51.9$\pm$2.0     & 4.5$\pm$0.2     \\
 CaII       &  8498.03          & 572.8$\pm$2.0    & 372.3$\pm$1.0        &   526.8$\pm$2.0     & 9.1$\pm$0.2     \\
 CaII       &  8542.09          & 670.5$\pm$2.0    & 430.7$\pm$1.0        &   636.5$\pm$2.0     & 8.0$\pm$0.2     \\
 CaII       &  8662.14          & 584.9$\pm$2.0    & 370.6$\pm$1.0        &   546.2$\pm$2.0     & 8.4$\pm$0.2     \\
 Pa12       &  8750.47          &  97.0$\pm$2.0    &  64.3$\pm$1.3        &    75.4$\pm$2.0     &   -             \\
 Pa9        &  9229.01          & 183.7$\pm$2.0    &  89.3$\pm$1.3        &   144.5$\pm$2.0     & 5.2$\pm$0.2     \\
\cline{1-6}
\enddata
\end{deluxetable}

\begin{deluxetable}{cccccc}
\tablewidth{0pt}
\tabletypesize{\footnotesize}
\tablecaption{\label{tab:tab6} Fluxes of lines detected with the Asiago telescopes.} 
\tablehead{
\colhead{Date}   &\multicolumn{5}{c}{Flux$^a\pm \Delta$ Flux (10$^{-15}$ erg s$^{-1}$ cm$^{-2}$) } 
}                  
\startdata      
                &  H$\alpha^b$  &   H$\beta$   & H$\gamma$  &  H$\delta$    & CaII H \\
\cline{1-6}
2015 Dec 23     &  1259$\pm$19   &   162$\pm$18  & 71$\pm$22  &  54$\pm$20  & 89$\pm$20\\
2015 Dec 27     &  1070$\pm$16   &   142$\pm$18  & 69$\pm$20  &  51$\pm$20  & 68$\pm$25\\
2015 Dec 29$^c$ &  1291$\pm$18   &   172$\pm$19  & 74$\pm$20  &  71$\pm$20  & 74$\pm$20\\
2016 Jan 21     &  1129$\pm$19   &   197$\pm$20  & 97$\pm$24  &  50$\pm$20  & 84$\pm$20\\  
2016 Feb 05$^d$ &  1076$\pm$12   &   177$\pm$12  & 58$\pm$17  &	-	        &  -       \\
2016 Feb 20     &  1051$\pm$18   &   118$\pm$18  & 64$\pm$20  &	-	        & 86$\pm$25\\
2016 Mar 13$^d$ &  1110$\pm$30   &   189$\pm$30  & 63$\pm$23  &	-	        & -        \\ 
2016 Mar 19     &  1056$\pm$18   &   115$\pm$18  & 52$\pm$20  &	-	        & -        \\
\cline{1-6}
\enddata
\tablenotetext{a}{Fluxes are the average of the values measured in the low- and high-resolution spectrum, unless specified otherwise.}
\tablenotetext{b}{Flux measured in the high-resolution spectrum (see text).}
\tablenotetext{c}{On this date, only a high-resolution spectrum is available.} 
\tablenotetext{d}{On this date, only a low-resolution spectrum is available.}
\end{deluxetable}


\begin{deluxetable}{ccccc}
\tablewidth{0pt}
\tabletypesize{\footnotesize}
\tablecaption{\label{tab:tab7} Fluxes of main lines detected with LUCI2.} 
\tablehead{
           &                   &  2016 Jan 25        &   2016 Mar 02               & 2016 Oct 04 
}                  
\startdata

Line        & $\lambda_{air}$   &   \multicolumn{3}{c}{Flux$\pm \Delta$ Flux}                            \\
            &  (\AA)            &   \multicolumn{3}{c}{ (10$^{-15}$ erg s$^{-1}$ cm$^{-2}$)}        \\
\cline{1-5}            
Pa$\delta$ & 10049.37          & 238.1$\pm$2.0        &   181.5$\pm$2.0            & 40.0$\pm$2.0 \\ 
HeI        & 10830.3           & (-32.2) 79.8$\pm$2.0 &   (-18.0) 113.8$\pm$2.0    & (-3.5) 7.2 $\pm$0.3 \\ 
Pa$\gamma$ & 10938.09          & 274.9$\pm$2.0        &   196.2$\pm$2.0            & 65.3$\pm$1.0 \\ 
NaI        & 11406.90          & 27.9 $\pm$3.0        &    32.6$\pm$3.0            & - \\  
Pa$\beta$  & 12818.08          & 362.5$\pm$2.0        &   282.8$\pm$2.0            & 71.8$\pm$1.0 \\ 
MgI  	   & 15029/15051       &  75.6$\pm$4.0        &    52.3$\pm$3.0            & - \\
Br19       & 15264.71          &  26.8$\pm$4.0        &    23.4$\pm$3.0            & - \\ 
Br18       & 15345.98          &  27.7$\pm$4.0        &    26.8$\pm$3.0            & - \\
Br17       & 15443.14          &  34.3$\pm$4.0        &    34.0$\pm$3.0            & - \\
Br16       & 15560.70          &  31.2$\pm$4.0        &    36.8$\pm$3.0            & - \\  
Br15       & 15704.95          &  41.0$\pm$4.0        &    37.4$\pm$3.0            & - \\
Br14+OI    & 15885/15892       &  111.4$\pm$4.0       &    97.3$\pm$3.0            & - \\ 
Br13       & 16113.71          &  51.1$\pm$4.0        &    47.6$\pm$3.0            & - \\
Br12       & 16411.67          &  72.0$\pm$4.0        &    72.0$\pm$3.0            & - \\
Br11       & 16811.11          &  73.4$\pm$4.0        &    75.1$\pm$3.0            & 15.6$\pm$3.0 \\
Br10       & 17366.85          &  84.9$\pm$4.0        &    80.0$\pm$3.0            & 10.1$\pm$3.0 \\\
H$_2$ 1-0 S(1)&   -            &         -            &      -                     & 3.9$\pm$0.7 \\
Br$\gamma$ & 21655.29          &  83.8$\pm$4.0        &    75.6$\pm$3.0            & 15.5$\pm$0.7 \\   
NaI        & 22062.4           &  17.5$\pm$4.0        &    10.0$\pm$3.0            & - \\
NaI        & 22089.7           &  10.9$\pm$4.0        &     9.9$\pm$3.0            & - \\
CO 2-0     & 22943             &  185.1$\pm$10.0      &    140.8$\pm$10.0          & - \\
CO 3-1     & 23235             &  161.7$\pm$10.0      &    117.9$\pm$10.0          & - \\     
\cline{1-5}
\enddata
\end{deluxetable}


\begin{deluxetable}{ccccc}
\tablewidth{0pt}
\tabletypesize{\footnotesize}
\tablecaption{\label{tab:tab8} Mass accretion rate.} 
\tablehead{
Phase           &\multicolumn{4}{c}{\macc\,(\msunyr)$^a$}             
}                                                       
\startdata
                  &   C-1.5		             & C-2.9	         	    & L-1.5		           &   L-2.9  		            \\
\cline{1-5}
Quiescence$^b$    &  3.1$\pm$1.0 10$^{-9}$   &  2.0$\pm$0.7 10$^{-8}$ & 3.5$\pm$2.5 10$^{-9}$  &  1.6$\pm$1.1 10$^{-8}$	    \\
Rising$^c$        &  1.1$\pm$0.4 10$^{-7}$   &  6.8$\pm$2.4 10$^{-7}$ & 4.2$\pm$0.8 10$^{-8}$  &  3.1$\pm$0.5 10$^{-7}$	    \\ 
Peak$^d$          &  3.2$\pm$1.1 10$^{-7}$   &  1.9$\pm$0.7 10$^{-6}$ & 2.0$\pm$1.0 10$^{-7}$  &  9.6$\pm$3.1 10$^{-7}$	    \\   
Declining$^e$     &    -    		   &		    -	              & 8.0$\pm$2.2 10$^{-8}$  &  1.4$\pm$0.5 10$^{-7}$	    \\ 
Post-outburst$^f$ &  3.0$\pm$0.4 10$^{-8}$   &  6.4$\pm$0.4 10$^{-8}$ & 2.6$\pm$0.9 10$^{-8}$  &  1.8$\pm$0.8 10$^{-7}$	    \\ 
\cline{1-5} 
\enddata 
\tablenotetext{a}{Column labels refer to the method (C:'continuum', L:'lines') and to the adopted A$_V$ (1.5 and 2.9 mag).}
\tablenotetext{b}{Derived from data of Paper\,I (taken on 2014).}
\tablenotetext{c}{Derived from data Paper\,II (taken in 2015 October).}
\tablenotetext{d}{Average of the determinations between 2015 December and 2016 March.}
\tablenotetext{e}{Measured on the LUCI2 spectrum taken in 2016 October.}
\tablenotetext{f}{Measured on the MODS spectrum taken in 2016 December.}
\end{deluxetable}

\begin{deluxetable}{cccccc}
\tablewidth{0pt}
\tabletypesize{\footnotesize}
\tablecaption{\label{tab:tab9} Statistics$^a$ of the optical permitted lines.} 
\tablehead{
Element  & IP$^b$       &\multicolumn{2}{c}{V1118 Ori}    &\multicolumn{2}{c}{EX Lup}
}                                                       
\startdata
         &              & max(E$_{up}$)$^b$ & N$_{lines}$    & max(E$_{up}$)$^b$ & N$_{lines}$ \\
\cline{1-6}             
Li I     & 5.39         &  1.8    &	  2    & -	 &  - \\
Na I     & 5.13         &  2.1    &	  2    & 4.9	 &  3 \\  
Mg I     & 7.64         &  5.1    &	  1    & 7.6	 & 10 \\   
Si I     & 8.15         &  5.1    &	  1    & 7.5	 &  7 \\
Si II    & 16.34        & 10.1    &	  2    & 14.8	 &  6 \\
Ca I     & 6.11         &  4.5    &	 15    & 5.7	 & 27 \\
Ca II    & 11.87        &  3.1    &	  2    & 10.0	 &  3 \\  
Ti I     & 6.82         &  6.0    &	 15    & 3.9	 &  6 \\
Ti II    & 13.57        &  5.6    &	 21    & 5.6	 & 43 \\
V I      & 6.74         &  5.4    &	 10    & 4.7	 &  4 \\ 
Cr I     & 6.76         &  5.0    &	 11    & 5.0	 & 31 \\ 
Cr II    & 16.49        &  6.8    &	  5    & 6.8	 & 12 \\
Fe I     & 7.87         &  6.7    &	 112   & 7.2	 & $>$ 400 \\
Fe II    & 16.18        &  5.9    &	  39   & 8.2	 & $\sim$ 70\\
Co I     & 7.86         &  4.0    &	  2    & 4.2	 & 9 \\
Ni I     & 7.63         &  4.1    &	  4    & 6.1	 & 17\\
\cline{1-6} 
\enddata 
\tablenotetext{a}{The comparison between V1118 Ori and EX Lup spectra has been made within the range 3300$-$7300 \AA\, covered with the echelle of the 1.82m Asiago telescope.}
\tablenotetext{b}{Units are eV.}
\end{deluxetable}


\begin{deluxetable}{ccccccc}
\tablewidth{0pt}
\tabletypesize{\footnotesize}
\tablecaption{\label{tab:tab10} Parameters of the main Gaussians fitting the \hi\, lines. Units are km s$^{-1}$.} 
\tablehead{
Date  &    \multicolumn{2}{c}{Narrow}   &   \multicolumn{2}{c}{Broad}        &   \multicolumn{2}{c}{P Cygni}           
}
\startdata
      &     Center       &  FWHM        &     Center  &  FWHM                  &     Center  &  FWHM                    \\
\cline{1-7}
\multicolumn{7}{c}{H$\alpha$}\\ 
\cline{1-7}
2015 Dec 23  &   -9   &  209  & -101 &   449	&  -195 &   42 \\
2015 Dec 27  &   -4   &  172  & -3   &   339	&  -191 &   41 \\
2015 Dec 29  &   -12  &  203  & -7   &   387	&  -195 &   38 \\
2016 Jan 21  &   -8   &  192  & -37  &   383	&  -213 &   60 \\
2016 Feb 20  &   -17  &  185  & -25  &   341	&  -243 &   71 \\
2016 Mar 19  &   -3   &  205  &  1   &   402	&  -259 &   61 \\
\cline{1-7}
\multicolumn{7}{c}{H$\beta$}\\ 
\cline{1-7}
2015 Dec 23  &    3	 & 197   & -111 &   488   &  -159 &  65  \\
2015 Dec 27  &   -4  & 174   & -1   &   434   &  -176 &  69  \\
2015 Dec 29  &   -5	 & 203   & -1   &   387   &  -177 &  83  \\
2016 Jan 21  &   -5	 & 182   & -2   &   451   &  -212 &  83  \\
2016 Feb 20  &   -5	 & 174   & -2   &   350   &  -212 &  83  \\
2016 Mar 19  &   -5	 & 172   & -2   &   373   &  -212 &  83  \\
\cline{1-7}
\multicolumn{7}{c}{Pa$\beta$}\\ 
\cline{1-7}
2016 Jan 10  &   -3	 & 112   &  -5 & 282     &   -         &  -          \\
2016 Apr 05  &    10 & 109   & -23 & 247     &   -         &  -          \\
\cline{1-7}
\multicolumn{7}{c}{Pa$\gamma$}\\ 
\cline{1-7}
2016 Jan 10  &   -14	 & 112   & 2  & 234  &   -         & -           \\
2016 Apr 05  &   -12	 & 109   & 2  & 199  &   -         & -           \\
\cline{1-7}
\multicolumn{7}{c}{Br10}\\ 
\cline{1-7}
2016 Jan 10  &   3	 & 81   & 6  & 177       &  -         & -            \\
\cline{1-7}
\multicolumn{7}{c}{Br11}\\ 
\cline{1-7}
2016 Jan 10  &  5	 & 72   & -2 &  183    &  -         & -            \\
2016 Apr 05  &  4	 & 78   & -1 &  171    &  -         & -            \\ 
\cline{1-7}
\multicolumn{7}{c}{Br12}\\ 
\cline{1-7}
2016 Jan 10  &  4	 & 86   &  3 &  273$^a$&  -         & -            \\
2016 Apr 05  &  4	 & 99   &  3 &  264$^a$&  -         & -            \\
\cline{1-7}
\multicolumn{7}{c}{Br13}\\ 
\cline{1-7}
2016 Jan 10  &  6	 & 72  & -4  & 192   &  -         & -            \\
\cline{1-7}
\enddata 
\tablenotetext{a}{Contaminated by Ti\,I and V\,I lines.}
\end{deluxetable}


\begin{deluxetable}{cc}
\tablewidth{0pt}
\tabletypesize{\footnotesize}
\tablecaption{\label{tab:tab11} 2015-2016 outburst parameters of V1118 Ori.} 
\tablehead{
Parameter           & Value                         
}
\startdata
\cline{1-2} 
$\Delta V^a$        & 4 mag                         \\     
Duration            & $\sim$ 1 year                 \\
Rising velocity     & $\la$ 0.008 mag/day           \\
Declining velocity  & $\sim$ 0.015 mag/day          \\
Periodicity         & hints                         \\
$<\Delta$[V-R]$>^a$ & $\sim$ 0.4 mag                \\
$<\Delta$[H-K]$>^a$ & $\sim$ 0.0 mag                \\
Peak \macc          & 0.1-1.2 10$^{-6}$ \msunyr     \\
Electron density    & $\sim$ 4 10$^{11}$ cm$^{-3}$  \\
Metals              & mostly neutrals               \\
$<$FHWM$>$ neutrals & $<$ 100 km s$^{-1}$           \\
$<$FHWM$>$ \hi\, lines & up to 450 km s$^{-1}$      \\
Wind?               & yes, variable P Cyg           \\
\cline{1-2}   
\enddata 
\tablenotetext{a}{Variation with respect to the quiescence value.}
\end{deluxetable}

\begin{deluxetable}{cccc}
\tablewidth{0pt}
\tabletypesize{\footnotesize}
\tablecaption{\label{tab:tab12} Lines observed with the REOSC Echelle at the 1.82m Asiago telescope, grouped according to element}
\tablehead{
\colhead{Line}  & \colhead{$\lambda_{air}$ }    & \colhead{A$_{ij}$}  & \colhead{E$_{up}$ } 
}
\startdata
         & (\AA)                    & (s$^{-1}$)  &   (eV)             \\
\cline{1-4}\\[-5pt]

H$\delta$ &4101.73                  & 9.72e+05  & 13.220              \\
H$\gamma$ &4340.46                  & 2.53e+06  & 13.054  	      \\ 
H$\beta$  &4861.28    	            & 8.41e+06  & 12.748  	      \\ 
H$\alpha$ &6562.57                  & 4.41e+07  & 12.087  	      \\
He I     &4471.47                  & 2.47e+07  & 23.736		     \\
He I     &5875.6     	           & 7.09e+07  & 23.073		     \\
He I     &6678.15		   & 6.38e+07  & 23.074     	     \\
He I     &7065.19		   & 1.55e+07  & 22.718 	     \\
Li I 	 &6707.76		   & 3.70e+07  &  1.847 	     \\
Li I 	 &6707.91		   & 3.70e+07  &  1.847 	     \\
O I      &6300.30		   & 5.63e-03  &  1.967 	     \\
O I      &6363.77		   & 1.82e-03  &  1.967 	     \\
NaI D	 &5889.95		   & 6.24e+07  &  2.104              \\
NaI D    &5895.92		   & 6.22e+07  &  2.102 	     \\
Mg I     &5183.60		   & 5.61e+07  &  5.107	             \\
Si I     &3905.52		   & 1.39e+07  &  5.082	    	     \\
Si II    &6347.10		   & 6.11e+07  & 10.073	    	     \\
Si II    &6371.36		   & 6.05e+07  & 10.066	    	     \\
Ca I     &4226.73		   & 2.19e+08  &  2.932	             \\   
Ca I     &4302.53		   & 1.43e+08  &  4.779	    	     \\
Ca I     &4434.96		   & 6.37e+07  &  4.680	    	     \\
Ca I     &4454.77		   & 8.38e+07  &  4.681	    	     \\
Ca I     &5602.85		   & 1.78e+07  &  4.735	    	     \\
Ca I     &6102.72		   & 9.53e+06  &  3.910	    	     \\

Ca I     &6122.22		   & 2.85e+07  &  3.910	    	     \\
Ca I     &6162.17		   & 4.74e+07  &  3.910	    	     \\
Ca I     &6169.06		   & 8.12e+06  &  4.532	    	     \\
Ca I     &6439.07		   & 4.26e+07  &  4.450	    	     \\
Ca I     &6449.81		   & 6.68e+06  &  4.443	    	     \\
Ca I     &6462.57		   & 3.75e+07  &  4.440	    	     \\
Ca I     &6471.66		   & 4.66e+06  &  4.440	    	     \\
Ca I     &6717.69		   & 1.08e+07  &  4.554	    	     \\
Ca I     &7148.15		   & 3.90e+07  &  4.443	    	     \\
CaII H   &3933.66		   & 1.40e+08  &  3.150	    	     \\
CaII K   &3968.46		   & 1.36e+08  &  3.123	    	     \\
Ti I     &4783.30		   & 6.20e+04  &  3.41	    	     \\
Ti I     &4899.91		   & -         &  4.41	    	     \\
Ti I]    &4934.58		   & -         &  4.84	    	     \\
Ti I     &5006.99		   & -         &  5.69	    	     \\
Ti I     &5136.64		   & -         &  5.96      	     \\
Ti I     &5153.48		   & -         &  5.69      	     \\
Ti I     &5265.71		   & -         &  4.48      	     \\
Ti I/V I &5473.50/5473.59          & -         &  4.59/4.62          \\
Ti I     &5527.61		   & -         &  4.67      	     \\
Ti I     &5711.89		   & -         &  4.47   	     \\
Ti I     &6359.20		   & -         &  4.10  	   \\
Ti I     &6583.64		   & -         &  5.20  	   \\
Ti I     &6738.33		   & -         &  5.30  	   \\
Ti I     &6743.12		   & 6.90e+05  &  2.74  	   \\
Ti I     &6772.66		   & -         &  6.05  	   \\  
Ti II    &4294.10		   & 4.70e+06  &  3.97  	   \\
Ti II    &4307.86		   & 4.60e+06  &  4.04  	   \\
Ti II    &4314.97		   & 1.30e+07  &  4.03  	   \\
Ti II    &4320.96		   & 2.40e+06  &  4.03  	   \\
Ti II    &4394.05		   & 2.20e+06  &  4.04  	   \\
Ti II    &4399.77		   & 3.10e+06  &  4.05  	   \\
Ti II    &4417.72		   & 2.10e+06  &  3.97  	   \\
Ti II    &4443.79		   & 1.10e+07  &  3.87               \\				 
Ti II    &4450.49		   & 2.00e+06  &  3.87               \\
Ti II    &4464.45		   & 7.00e+05  &  3.94               \\
Ti II    &4468.50		   & 1.00e+07  &  3.90       	     \\
Ti II    &4501.27		   & 9.80e+06  &  3.86      	     \\
Ti II    &4563.77		   & 8.80e+06  &  3.94               \\
Ti II    &4571.98		   & 1.20e+07  &  4.28         	     \\
Ti II/Fe II&4629.28/.34 &2.20e+05/1.30e+05  & 3.86/5.48              \\
Ti II    &4779.98		   & 6.20e+06  &  4.64      	     \\
Ti II    &4798.52		   & 1.80e+05  &  3.66      	     \\
Ti II    &4805.08		   & 1.10e+07  &  4.64      	     \\
Ti II    &4911.19		   & 3.20e+07  &  5.65               \\
Ti II    &5129.15		   & 1.00e+06  &  4.31               \\
Ti II    &5188.68		   & 2.50e+06  &  3.97	             \\
Ti II    &5336.77		   & 5.80e+05  &  3.90	    	     \\
Ti II    &5381.01		   & 3.20e+05  &  3.87	    	     \\
V I      &3912.88		   & 3.30e+06  &  4.23	             \\
V I      &3914.39		   & -         &  5.44	    	     \\
V I      &4405.01		   & 3.40e+05  &  3.09      	     \\
V I      &4753.93		   & 1.50e+07  &  4.68               \\
V I      &5410.46		   & -         &  4.86      	     \\
V I]     &5413.36		   & -         &  2.56      	     \\
V I      &5415.24		   & 3.10e+07  &  4.66      	     \\
V I      &5463.34		   & -         &  4.85      	     \\
V I      &5489.93                  & 2.9e+07   &  4.63	             \\       
V I      &5762.73/.83	           & -/-       &  4.72/4.28          \\
V  I     &7005.45		   & -         &  3.89               \\
Cr I     &4254.33		   & 3.15e+07  &  2.91               \\
Cr I     &4580.04		   & 2.40e+06  &  3.65               \\
Cr I     &4903.22		   & 7.40e+06  &  5.07               \\
Cr I     &5204.51		   & 5.09e+07  &  3.32               \\
Cr I     &5206.04		   & 5.14e+07  &  3.32               \\
Cr I     &5208.44		   & 5.06e+07  &  3.32               \\
Cr I     &5247.57		   & 1.90e+06  &  3.32               \\  
Cr I     &5298.27		   & 3.30e+06  &  3.32               \\
Cr I     &5345.81		   & 4.90e+06  &  3.32               \\
Cr I     &5348.61		   & 1.70e+06  &  3.32               \\
Cr I     &5409.77		   & 6.20e+06  &  3.32               \\
Cr II    &4558.66		   & 8.80e+06  &  6.79               \\
Cr II    &4634.07		   & 8.90e+06  &  6.75               \\
Cr II    &4824.13	           & 1.70e+06  &  6.44               \\
Cr II    &4848.23		   & 2.60e+06  &  6.42      	     \\
Cr II    &4876.47		   & 1.60e+06  &  6.41      	     \\
Fe I     &4063.59		   & 6.80e+07  &  4.61      	     \\
Fe I     &4118.54		   & 5.80e+07  &  6.58               \\
Fe I     &4123.72/.76              & 2.79/4.50e+05/ &  6.00/5.6      \\
Fe I     &4132.05		   & 1.20e+07  &  4.61      	     \\
Fe I     &4154.49		   & 4.96e+07  &  5.82               \\
Fe I     &4173.31                  & 2.10e+06  &  5.82               \\
Fe I     &4191.43		   & 3.19e+07  &  5.43               \\
Fe I     &4198.63		   & 2.46e+07  &  6.37      	     \\
Fe I     &4202.02		   & 8.22e+06  &  4.43      	     \\
Fe I     &4216.18		   & 1.84e+04  &  2.93               \\
Fe I     &4233.60		   & 2.33e+07  &  5.41      	     \\  
Fe I     &4250.11/.79	           &2.63/1.00e+07&  5.39/4.47        \\   
Fe I     &4260.47                  & 4.67e+07  &  6.31/5.3           \\  
Fe I     &4271.15/.76              & 2.26/2.28e+07&  5.35/4.3        \\
Fe I     &4282.40		   & 1.10e+07  &  5.07               \\
Fe I     &4289.91		   & 2.69e+06  &  6.29               \\
Fe I     &4324.94                  & 1.63e+04  &  5.06               \\     
Fe I     &4375.98                  & 2.96e+05  &  5.8 	             \\
Fe I     &4415.12		   & 1.19e+07  &  4.42	             \\
Fe I     &4427.29                  & 6.87e+06  &  6.45	   	     \\
Fe I     &4442.34		   & 3.76e+06  &  4.99	    	     \\
Fe I     &4443.19		   & 1.26e+07  &  5.64      	     \\
Fe I     &4447.71		   & 5.11e+06  &  5.01	    	     \\
Fe I     &4459.11		   & 2.52e+06  &  4.96	             \\
Fe I     &4461.65		   & 2.95e+04  &  2.87	       	     \\
Fe I     &4466.55		   & 1.58e+07  &  5.61	    	     \\
Fe I     &4476.01		   & 1.36e+07  &  5.61	    	     \\
Fe I     &4482.25		   & 1.95e+06  &  4.98	    	     \\
Fe I     &4489.73		   & 1.19e+04  &  2.88	             \\
Fe I     &4494.56		   & 3.45e+06  &  4.95	             \\
Fe I     &4528.61		   & 5.44e+06  &  4.91	             \\
Fe I/Ti II&4533.96/.97     & 2.21e+04/9.20e+06 & 5.68/3.97           \\
Fe I     &4583.72		   & 5.90e+05  &  5.82	    	     \\
Fe I     &4602.94		   & 1.48e+05  &  4.18	    	     \\
Fe I/Cr II&4618.76/.80     & 1.60e+05/6.1e+06  &  5.63/6.76   	     \\
Fe I     &4654.60/.63      & 9.98e+06/9.33e+05 &6.26/5.87 	     \\
Fe I     &4667.45		   & 4.18e+06  &  6.25	             \\
Fe I     &4678.84		   & 2.23e+07  &  6.25	             \\
Fe I     &4702.91		   & 3.92e+04  &  5.87	    	     \\
Fe I     &4736.77		   & 4.90e+06  &  5.83	    	     \\
Fe I     &4786.80		   & 1.21e+06  &  5.61	             \\
Fe I     &4871.31		   & 3.22e+07  &  5.41	    	     \\
Fe I     &4877.60		   & 5.79e+04  &  5.53	    	     \\
Fe I     &4891.49		   & 4.10e+07  &  5.39	    	     \\
Fe I     &4918.99		   & 2.31e+07  &  5.39	    	     \\
Fe I     &4920.50		   & 4.90e+07  &  5.35	             \\
Fe I     &4939.24                  & 5.78e+06  &  6.66               \\
Fe I     &4957.60                  & 5.53e+07  &  5.31	             \\
Fe I     &4985.25		   & 2.96e+07  &  6.41	             \\		      
Fe I     &4994.12		   & 3.18e+04  &  3.39	             \\
Fe I     &5006.11		   & 7.28e+06  &  5.31	             \\
Fe I     &5012.06		   & 5.50e+04  &  3.33	             \\				  
Fe I     &5030.77		   & 2.60e+04  &  5.70	             \\	  
Fe I     &5041.07                  & 2.15e+05  &  3.42               \\ 
Fe I     &5049.81		   & 2.31e+06  &  4.73	             \\ 
Fe I     &5051.63		   & 4.65e+04  &  3.37               \\
Fe I     & 5064.95	           & 2.88e+06  &  6.70               \\
Fe I     & 5068.76		   & 3.37e+06  & 5.39                \\
Fe I     & 5079.22/.74	 & 7.39e+05/5.19e+04   & 4.64/3.43           \\
Fe I     & 5083.33		   & 4.06e+04  & 3.40                \\
Fe I     & 5098.57/.69	 & 6.08e+05/2.97e+05   & 6.36/4.61           \\
Fe I     & 5107.64		   & 1.95e+05  & 3.98                 \\
Fe I     & 5110.35		   & 1.26e+06  & 6.0		      \\
Fe I	 & 5123.72	           & 7.24e+04  & 3.4		      \\
Fe I     & 5127.36	           & 1.14e+04  & 3.3		      \\
Fe I     & 5139.25	           & 1.13e+07  & 5.4		     \\
Fe I     & 5141.73	           & 7.62e+05  & 4.8		     \\
Fe I     & 5150.83	           & 3.10e+04  & 3.4		     \\
Fe I     & 5167.49		   & 2.00e+06  & 3.88   	     \\
Fe I     & 5171.60		   & 4.46e+05  & 3.8		     \\
Fe I     & 5191.45		   & 2.97e+07  & 5.43   	     \\
Fe I     & 5194.94		   & 2.87e+05  & 3.94		     \\
Fe I     & 5202.25		   & 5.14e+06  & 6.64		     \\
Fe I     & 5216.27		   & 3.47e+05  & 3.98		     \\
Fe I     & 5227.15/.19	           & 1.40e+06/4.23e+06	& 4.80/3.93  \\      
Fe I     & 5232.94		   & 2.57e+07  & 5.31		     \\
Fe I     & 5250.64		   & 3.10e+05  & 4.56		     \\
Fe I     & 5254.97		   & 2.15e+04  & 6.65		     \\
Fe I     & 5270.35		   & 2.50e+06  & 3.96		     \\
Fe I     & 5302.30		   & 1.17e+06  & 5.62		     \\ 
Fe I     & 5307.36		   & 3.49e+04  & 3.94		     \\
Fe I     & 5324.17		   & 2.74e+07  & 5.54		     \\
Fe I     & 5328.03		   & 1.15e+06  & 3.24		     \\
Fe I     & 5332.89		   & 3.00e+04  & 3.88		     \\
Fe I     & 5341.02		   & 4.10e+05  & 3.93		     \\
Fe I     & 5371.48		   & 1.05e+06  & 3.27		     \\
Fe I     & 5383.36		   & 5.60e+07  & 6.61		     \\
Fe I     & 5393.16		   & 6.34e+06  & 5.54		     \\
Fe I     & 5397.12		   & 2.59e+05  & 3.21		     \\
Fe I     & 5405.77		   & 1.09e+06  & 3.28		     \\
Fe I     & 5429.69		   & 4.27e+05  & 3.24		     \\
Fe I     & 5434.52		   & 1.71e+06  & 3.29		     \\
Fe I     & 5446.91		   & 5.30e+05  & 3.26		     \\
Fe I     & 5455.60		   & 1.02e+06  & 3.28		     \\
Fe I     & 5487.74		   & 9.70e+06  & 6.58		     \\
Fe I     & 5497.51		   & 6.25e+04  & 3.27		     \\
Fe I     & 5501.46		   & 2.70e+04  & 3.21		     \\
Fe I     & 5506.77		   & 5.01e+04  & 3.24		     \\
Fe I     & 5572.84 		   & 3.55e+07  & 5.62      	     \\
Fe I     & 5586.75		   & 3.81e+07  & 5.59		     \\
Fe I     & 5615.64		   & 4.02e+07  & 5.5		     \\
Fe I     & 5624.54		   & 1.17e+07  & 5.6		     \\
Fe I     & 5658.81		   & 7.64e+06  & 5.5		     \\
Fe I     & 5662.51		   & 7.83e+06  & 6.3		     \\
Fe I     & 5701.54		   & 3.78e+05  & 4.7		     \\
Fe I     & 5709.37		   & 3.41e+06  & 5.5		     \\
Fe I     & 5753.12		   & 1.72e+07  & 6.4		     \\
Fe I     & 5755.34	           & 6.96e+04  & 5.79       	     \\
Fe I     & 5791.01		   & 1.12e+05  & 5.3		     \\
Fe I     & 5853.68		   & 1.67e+05  & 6.31		     \\
Fe I     & 5914.11/.20	   & 1.61e+07/1.57e+07 & 6.70/6.70  	     \\ 
Fe I     & 5930.18		   & 1.60e+07  & 6.74		     \\
Fe I     & 5934.65		   & 1.80e+06  & 6.0		     \\
Fe I     & 6003.01		   & 1.60e+06  & 5.95		     \\
Fe I     & 6007.95/8.55	 & 6.68E+06/2.11e+06   & 6.72/5.95	     \\
Fe I     & 6024.05		   & 1.30e+07  & 6.61		     \\
Fe I     & 6027.05		   & 1.00+06   & 6.13		     \\
Fe I     & 6065.48		   & 1.18e+06  & 4.65		     \\
Fe I     & 6136.61/7.69	 & 1.01e+06/1.14e+06   & 4.47/4.61	     \\ 
Fe I     & 6141.73		   & 1.60e+06  & 5.62		     \\
Fe I     & 6173.33		   & 2.31e+05  & 4.23	   	     \\
Fe I     & 6191.55		   & 4.90e+05  & 4.43		     \\
Fe I     & 6200.31		   & 9.24e+04  & 4.61		     \\
Fe I     & 6213.43		   & 1.30e+05  & 4.22		     \\
Fe I     & 6219.28		   & 1.27e+05  & 4.19   	     \\
Fe I     & 6230.72		   & 1.15e+06  & 4.55		     \\
Fe I     & 6246.31		   & 6.40e+06  & 3.88		     \\
Fe I     & 6252.55		   & 3.19e+05  & 4.39		     \\
Fe I     & 6254.25		   & 2.14e+05  & 4.26		     \\
Fe I/Ni I& 6256.36/.36   & 4.50e+04/1.90e+05   & 4.43/3.65  	     \\
Fe I     & 6265.13		   & 6.84e+04  & 4.15		     \\
Fe I     & 6270.22		   & 1.10e+05  & 4.83		     \\
Fe I     & 6280.77		   & 1.23e+06  & 6.55		     \\
Fe I     & 6297.79		   & 6.12e+04  & 4.19		     \\
Fe I     & 6301.50		   & 1.03e+07  & 5.62		     \\
Fe I     & 6318.01		   & 1.31e+05  & 4.42		     \\
Fe I     & 6322.68	           & 6.87e+04  & 4.55		     \\
Fe I	 & 6335.33		   & 1.40e+05  & 4.15		     \\
Fe I     & 6355.02		   & 1.30e+05  & 4.79		     \\
Fe I     & 6380.74		   & 1.30e+06  & 6.13		     \\
Fe I     & 6393.60		   & 4.40e+05  & 4.37		     \\
Fe I     & 6400.00		   & 1.48e+07  & 5.54		     \\
Fe I     & 6408.01		   & 5.16e+06  & 5.62		     \\ 
Fe I     & 6411.64		   & 9.71e+06  & 5.59		     \\
Fe I/Fe II& 6416.93/.92  & 2.52e+06/3.80E+04   & 6.73/5.82 	     \\
Fe I     & 6421.35		   & 3.49e+05  & 4.21		     \\
Fe I     & 6430.84		   & 1.77e+05  & 4.10	   	     \\
Fe I     & 6475.62		   & 5.13e+04  & 4.47		     \\
Fe I     & 6481.87		   & 3.52e+03  & 4.19		     \\
Fe I     & 6494.98		   & 7.67e+05  & 4.31		     \\
Fe I     & 6546.23		   & 7.00e+05  & 4.65		     \\
Fe I     & 6593.87  		   & 5.28e+04  & 4.31		     \\ 
Fe I	 & 6604.58	           & 7.74e+05  & 6.71		     \\
Fe I     & 6609.11		   & 4.26e+04  & 4.43   	     \\
Fe I     & 6663.44		   & 5.72e+05  & 4.28		     \\
Fe I     & 6750.15		   & 1.40e+05  & 4.26		     \\
Fe I	 & 6945.20		   & 1.18e+05  & 4.21		     \\
Fe I     & 6978.85		   & 1.76e+05  & 4.26 	  	     \\
Fe I     & 6999.88	           & 4.20e+05  & 5.87		     \\
Fe I     & 7024.06		   & 1.20e+05  & 5.84		     \\
Fe I     & 7164.44		   & 7.98e+06  & 5.92	 	     \\
Fe I     & 7207.38	           & 9.04e+06  & 5.87      	     \\
Fe I     & 7307.93	           & 1.45e+06  & 5.84       	     \\
Fe II    & 4173.45	           & 4.20e+05  & 5.55		     \\
Fe II    & 4178.86	           & 1.60e+05  & 5.55   	     \\
Fe II    & 4296.57	           & 5.90e+04  & 5.58	             \\
Fe II    & 4303.17  		   & 2.90e+05  & 5.58		     \\
Fe II    & 4351.76		   & 4.86e+05  & 5.55	   	     \\
Fe II/Fe I& 4385.38/5.25   & 4.70e+05/1.08e+04 & 5.60/5.84           \\
Fe II    & 4472.92	    	   & 3.10e+04  & 5.61       	     \\
Fe II    & 4491.40  		   & 1.70e+05  & 5.61		     \\
Fe II    & 4508.28	           & 1.00e+06  & 5.60	             \\ 
Fe II    & 4515.33		   & 1.80e+05  & 5.59	     	     \\ 
Fe II    & 4520.22  		   & 1.00e+05  & 5.55		     \\ 
Fe II    & 4522.63		   & 7.60e+05  & 5.58	             \\
Fe II    & 4541.52		   & 7.20e+04  & 5.58       	     \\
Fe II    & 4549.47		   & 9.60e+05  & 5.55		     \\
Fe II    & 4555.89		   & 2.10e+05  & 5.55		     \\ 
Fe II    & 4576.34		   & 4.80e+04  & 5.55		     \\
Fe II    & 4620.52		   & 2.00e+04  & 5.51		     \\
Fe II    & 4656.98		   & 1.20e+04  & 5.55		     \\
Fe II    & 4670.18		   & 3.00e+03  & 5.24		     \\
Fe II    & 4731.45		   & 1.60e+04  & 5.51		     \\
Fe II    & 4923.92		   & 3.30e+06  & 5.41		     \\
Fe II    & 5018.43		   & 2.70e+06  & 5.36		     \\
Fe II    & 5197.57		   & 4.90e+05  & 5.62		     \\
Fe II    & 5234.62		   & 3.60e+05  & 5.59		     \\
Fe II    & 5276.00		   & 3.40e+05  & 5.55		     \\
Fe II    & 5284.10		   & 1.90e+04  & 5.24		     \\
Fe II    & 5316.61		   & 3.30e+05  & 5.48		     \\
Fe II    & 5325.55		   & 7.40e+04  & 5.55		     \\
Fe II    & 5362.86		   & -	       & 5.51		     \\
Fe II    & 5425.25		   & 9.90e+03  & 5.48	  	     \\
Fe II/Ti I  & 5534.84		   & 2.60e+04/-& 5.48/3.13	     \\
Fe II    & 5991.37		   & 3.40e+03  & 5.22		     \\
Fe II    & 6084.11		   & 2.40e+03  & 5.24		     \\
Fe II    & 6147.73		   & 1.3e+05   & 5.90		     \\
Fe II    & 6238.37		   & 7.5e+04   & 5.88		     \\
Fe II    & 6247.56		   & 1.30e+05  & 5.87		     \\
Fe II    & 6432.68		   & 4.90e+03  & 4.82		     \\
Fe II    & 6456.37		   & 1.30e+05  & 5.82		     \\
Fe II    & 6516.05  		   & 7.00e+03  & 4.79	   	     \\
Co I     & 6128.23/9.09    & 5.06e+05/3.60e+04 & 5.69/4.04	     \\
Co I     & 7016.62		   & 4.60e+05  & 3.77		     \\
Ni I     & 5476.91		   & 9.50e+06  & 4.09		     \\
Ni I     & 6108.12		   & 1.30e+05  & 3.71		     \\
Ni I     & 6643.64		   & 1.50e+05  & 3.54		     \\
Ni I     & 6767.77	           & 3.30e+05  & 3.66	   	     \\
INDEF    & 4990.7	           &  -	       &  -     	     \\
INDEF    & 5598.33  		   &  -        &  -		     \\ 
INDEF    & 5956.69		   &  -        &  -     	     \\
\cline{1-4}							    
\enddata						
\end{deluxetable}

\begin{deluxetable}{cccc}
\tablewidth{0pt}
\tabletypesize{\footnotesize}
\tablecaption{\label{tab:tab13} Lines observed with TNG/GIANO, grouped according to element}
\tablehead{
\colhead{Line}  & \colhead{$\lambda_{vac}$ }    & \colhead{A$_{ij}$}  & \colhead{E$_{up}$ } 
}
\startdata
                & (nm)                          &     (s$^{-1}$)      &   (eV)               \\
\cline{1-4}\\[-5pt]
H I Pa$\delta$  & 1005.21 & 3.35e+05  & 13.321	 \\    
H I Pa$\gamma$  & 1094.10 & 7.77e+05  & 13.221	 \\    
H I Pa$\beta$   & 1282.12 & 2.20e+06  & 13.054	 \\    
H I Br20       & 1519.59 & 1.18e+03  & 13.564   \\    
H I Br19       & 1526.47 & 1.53e+03  & 13.560   \\    
H I Br18       & 1534.59 & 2.01e+03  & 13.556   \\    
H I Br17       & 1544.31 & 2.69e+03  & 13.551   \\    
H I Br16       & 1556.06 & 3.67e+03  & 13.545   \\    
H I Br15       & 1570.49 & 5.11e+03  & 13.538   \\    
H I Br14       & 1588.48 & 7.28e+03  & 13.529   \\    
H I Br13       & 1611.37 & 1.07e+04  & 13.518   \\    
H I Br12       & 1641.16 & 1.62e+04  & 13.504   \\    
H I Br11       & 1681.11 & 2.55e+04  & 13.486	\\    
H I Br10       & 1736.68 & 4.23e+04  & 13.462   \\    
He I           & 1083.3  & 1.02e+07  & 20.964	\\    
C I            & 1032.91 & 6.56e+04  & 10.531	\\    
C I            & 1068.60 & 1.25e+07  &  8.643	\\    
C I            & 1068.82 & 9.29e+06  &  8.640	\\    
C I            & 1069.41 & 1.67e+07  &  8.647	\\    
C I            & 1071.02 & 6.92e+06  &  8.640	\\    
C I            & 1073.24 & 4.13e+06  &  8.643	\\    
C I            & 1166.20 & 8.85e+06  &  9.834	\\    
C I            & 1175.80 & 2.01e+07  &  9.697	\\    
C I            & 1689.50 & 1.14e+07  &  9.736	\\    
O I            & 1128.94 & 2.12e+07  & 12.087	\\    
O I            & 1128.95 & 1.18e+07  & 12.087	\\    
O I            & 1129.00 & 3.11e+07  & 12.087   \\    
Na I           & 2208.97 & 6.51e+06  &  3.752   \\    
Mg I           & 1081.40 & 1.92e+07  &  7.092   \\    
Mg I           & 1183.14 & 2.22e+07  &  5.394   \\    
Mg I           & 1208.69 & 1.54e+07  &  6.779	\\   
Mg I           & 1502.90 & 1.35e+07  &  5.933   \\    
Mg I           & 1504.43 & 1.34e+07  &  5.932   \\    
Mg I           & 1575.32 & 1.15e+07  &  6.719   \\    
Mg I           & 1577.00 & 1.23e+07  &  6.719   \\   
Mg I           & 1711.33 & 8.86e+06  &  6.118   \\   
Al I           & 1312.70 & 1.40e+07  &  4.087   \\   
Al I           & 1315.43 & 1.39e+07  &  4.085   \\   
Si I           & 1037.41 & 3.96e+06  &  6.124   \\   
Si I           & 1060.63 & 5.10e+06  &  6.098   \\   
Si I           & 1066.38 & 9.11e+06  &  6.083   \\   
Si I           & 1075.23 & 1.01e+07  &  6.083   \\   
Si I           & 1078.98 & 2.45e+07  &  6.078   \\
Si I	       & 1098.23 & 5.03e+06  &  6.082	\\
Si I	       & 1161.15 & 1.07e+05  &  7.290	\\
Si I	       & 1195.35 &      -    &  8.077	\\
Si I	       & 1198.74 & 1.40e+07  &  5.964	\\
Si I	       & 1199.48 & 1.04e+07  &  5.954	\\  
Si I	       & 1203.47 & 1.72e+07  &  5.984	\\
Si I	       & 1210.68 & 6.11e+06  &  5.954	\\
Si I	       & 1227.43 & 3.32e+06  &  5.964	\\
Si I	       & 1589.27 & 8.67e+06  &  5.862	\\
Si I	       & 1596.44 & 8.25e+06  &  6.761	\\
Si I	       & 1668.53 & 2.61e+06  &  6.727	\\
Ca I	       &  999.88 & 2.54e+06  &  5.983	\\
Ca I	       & 1000.02 & 4.44e+05  &  5.983	\\
Ca I	       & 1045.99 & 1.65e+06  &  5.920	\\
Ca I	       & 1050.50 &  -	     &  6.056   \\
Ca I	       & 1166.37 &  -	     &  6.107   \\
Ca I	       & 1169.41 &  -	     &  6.109   \\
Ca I	       & 1203.50 &  -	     &  6.078   \\
Ca I	       & 1210.91 &  1.06e+07 &  5.578   \\
Ca I	       & 1577.40 &  -	     &  6.107   \\
Ti I	       & 1039.96 & 7.1e+04   &  2.040   \\
Ti I	       & 1049.92 & 6.7e+04   &  2.016   \\
Ti I	       & 1058.93 &  -	     &  1.996   \\
Ti I	       & 1184.33 &  -	     &  2.344   \\
Ti I	       & 1533.94 & 2.84e+05  &  2.695   \\
Ti I	       & 1638.57 &  -        &  3.089   \\
Fe I	       & 1142.54 & 3.75e+04  &  3.283	\\
Fe I	       & 1159.67 & 1.85e+05  &  3.291	\\    
Fe I	       & 1161.07 & 9.85e+04  &  3.265	\\ 
Fe I	       & 1164.14 & 4.25e+04  &  3.240	\\
Fe I	       & 1169.31 & 1.45e+05  &  3.283	\\
Fe I	       & 1178.64 & 2.15e+05  &  3.883	\\ 
Fe I	       & 1188.60 & 1.35e+05  &  3.240	\\
Fe I	       & 1188.73 & 7.53e+04  &  3.265	\\ 
Fe I	       & 1197.63 & 1.75e+05  &  3.211	\\ 
Fe I	       & 1529.87 & 1.36e+07  &  6.119   \\ 
H$_2$	        & 2121.82 & 3.47e-07  &  0.861   \\
Na I	       & 22062.4 & 6.54E+06  &  3.753   \\ 
Na I	       & 22089.7 & 6.51E+06  &  3.752   \\	 
CO 3-1         & 2222.84 &	  -  &  0.52    \\   
\cline{1-4}
\enddata
\end{deluxetable}

\end{document}